\documentstyle[prc,aps,preprint,psfig]{revtex}
\draft
\begin{document}

\title{$e^+e^-$ production in $p A$ reactions at SIS energies
\thanks{Work supported by GSI and BMBF} }
\author{E. L. Bratkovskaya  \\[2mm]
{\normalsize Institut f\"{u}r Theoretische Physik, Universit\"{a}t Giessen}\\
{\normalsize 35392 Giessen, Germany}}
\maketitle

\begin{abstract}
Detailed predictions for dilepton production from $p A$ reactions at
SIS energies are presented within a semi-classical BUU transport model
that includes the off-shell propagation of vector mesons
nonperturbatively and calculates the width of the vector mesons dynamically.
Different scenarios of in-medium modifications of vector mesons, such
as collisional broadening and dropping vector meson masses, are
investigated and the possibilities for an experimental observation of
in-medium effects in $p A$ reactions at 1--4 GeV are discussed for a
variety of nuclear targets.
\end{abstract}

\vspace{0.5cm}\noindent

\vspace{0.5cm}\noindent

\narrowtext
\newpage
\section{Introduction}
The modification of hadron properties  in the nuclear matter are of
fundamental interest
(cf. Refs. \cite{BrownRho,Shakin94,Klingl96,H&L92,Asakawa93}) as
QCD sum rules \cite{H&L92,Asakawa93,Leupold} as well as QCD inspired
effective Lagrangian models
\cite{BrownRho,Shakin94,Klingl96,Herrmann,asakawa,Chanfray,Rapp,Friman,RappNPA,Peters}
predict  significant changes of the vector mesons ($\rho$, $\omega$ and
$\phi$) with the nuclear density.  A more direct evidence for the
modification of vector mesons has been provided by the enhanced
production of lepton pairs above known sources in nucleus-nucleus
collisions at SPS energies \cite{CERES,Ullrich,HELIOS,HELI2}.  As
proposed by Li, Ko, and Brown \cite{Li} and Ko et al. \cite{Li96}, the
observed enhancement  in the invariant mass range $0.3 \leq M \leq 0.7$
GeV might be due to a shift of the $\rho$-meson mass following
Brown/Rho scaling \cite{BrownRho} or the Hatsuda and Lee sum rule
prediction~\cite{H&L92}.  The microscopic transport studies in Refs.
\cite{Cass95C,Cass96H,Brat97,CBRep98,Ernst} for these systems support these
results \cite{Li,Li96,Ko93,Ko95}.  However, also more conventional approaches
that describe a melting of the $\rho$-meson in-medium due to the strong
hadronic coupling (along the lines of
Refs.~\cite{Herrmann,asakawa,Chanfray,Rapp,Peters}) were found to be
compatible with the CERES data~\cite{Rapp,Cass95C,CBRW97}.

An alternative way to provide independent information about the hadron
properties in the medium is to use more elementary probes such a pions,
protons or photons as incoming particles. In such reactions the nuclear
matter is close to the ground state, i.e. at normal nuclear density,
however, in-medium effects  might be still significant to be observed
experimentally.

In this paper, therefore, the study of dilepton production from
heavy-ion, pion-nucleus collisions (cf. \cite{CBRep98,Effe_piA}) and
photon-nucleus reactions \cite{Effe99gam} is extended to proton-nucleus
reactions. Whereas earlier studies on $p+A$ reactions have involved a
perturbative scheme for vector meson production and their dileptonic
decay \cite{Wolf90,BCMas96} the latter processes will be evaluated
nonperturbatively on the basis of the resonance model
\cite{Effe99gam,EffePhD} in this work and the collisional width will be
calculated dynamically as a function of nucleon density, mass and
momentum of the vector mesons.  Furthermore, the off-shell propagation
of the vector mesons -- adopted from Refs. \cite{Cass_off1,Cass_off2}
-- will be included consistently.

One might argue that $p+A$ reactions should yield similar dilepton
spectra than $\pi$ or $\gamma$ induced reactions on nuclear targets.
However, it is not clear if the sensitivity to vector meson in-medium
effects is the same in $\gamma +A, \pi +A$ or $p+A$ reactions since in
$\gamma+A$ reactions the nucleus is illuminated rather uniformly while
in $\pi, p$ induced reactions the initial reaction happens close to the
surface. Note, that the wavelength of the impinging hadron is short
compared to the distance between nucleons in the target such that such
classical considerations may be employed.  Furthermore, in $\pi+A$
reactions initially high mass resonances are excited in the $\pi N$
reaction whereas in $p+A$ reactions the excitation of high mass
resonances -- which have some $\rho$ meson decay width -- is suppressed
since the energy is shared between two nucleons and a substantial
longitudinal kinetic energy is left. Thus it is not obvious that all
reactions finally result in similar dilepton spectra. On the other
hand, it is of importance that all reactions (including $A+A$) are
calculated within the same approach.

The analysis of dilepton production from $p+A$ reactions is of special
interest with respect to the future
dilepton experimental program of the HADES Collaboration at GSI.  The
detailed microscopic calculations on the basis of the resonance
Boltzmann-Uehling-Uhlenbeck (BUU) model \cite{Effe99gam,EffePhD} will
thus be performed for dilepton production in the systems to be measured
experimentally, i.e. $p C$, $p Ca$,  $p Pb$ at proton energies
of $E = 1.0, \ 2.0, \ 3.0$ and 4.0~GeV.

In these calculations different scenarios of in-medium modifications of
vector mesons such as the 'dropping mass' scenario -- following Brown/Rho
scaling \cite{BrownRho} -- or the Hatsuda and Lee sum-rule
prediction~\cite{H&L92} as well as the effect of collisional
broadening (cf. e.g. \cite{Boresk96}) will be employed
including, however, a dynamical width of the vector mesons that is
calculated dynamically and consistent with the vector meson
production/absorption amplitudes (or probabilities).

The paper is organized as follows:  In Section~2 the underlying
resonance model, that enters the coupled-channel BUU transport approach
is presented.  In Section~3 detailed predictions for these reactions
are given employing a high mass resolution for the dilepton pair
of $\Delta M=10$~MeV in view of
upcoming experiments with the HADES detector at GSI Darmstadt. Section
4 contains a summary and discussion of open problems.

\section{Description of the model}

\subsection{The resonance approach}

The analysis of dilepton production from $p C$, $p Ca$
and $p Pb$ collisions is performed within the resonance approach of
Refs. \cite{Effe99gam,EffePhD}.
This model is based on the resonance concept of nucleon-nucleon and
meson-nucleon interactions at low invariant energy $\sqrt{s} \ $
\cite{TeisZP97} by adopting all resonance parameters from the Manley
analysis \cite{Manley};
all states with at least 2 stars in Ref.~\cite{Manley}
are taken into account:
$P_{33}$(1232), $P_{11}$(1440), $D_{13}$(1520), $S_{11}$(1535),
$P_{33}$(1600), $S_{31}$(1620), $S_{11}$(1650), $D_{15}$(1675),
$F_{15}$(1680), $P_{13}$(1879), $S_{31}$(1900), $F_{35}$(1905),
$P_{31}$(1910), $D_{35}$(1930), $F_{37}$(1950), $F_{17}$(1990),
$G_{17}$(2190), $D_{35}$(2350). These resonances couple to the following
channels: $N \pi$, $N \eta$, $N \omega$, $\Lambda K$, $\Delta(1232)
\pi$, $N \rho$, $N \sigma$, $N(1440) \pi$, $\Delta(1232) \rho$
with respect to the production and decay.

It has been shown in Ref. \cite{TeisZP97} that the resonance model
provides a good description of the experimental data on one- and
two-pion production in nucleon-nucleon collisions at low energy.
However, with increasing bombarding energy the resonance contributions
underestimate the data; the missing yield is then treated as a
background term to the resonance amplitude.  This background term
'mimics' $t$-channel particle production mechanism as well as other
non-resonance contributions (e.g., direct $NN\to NN\pi$, without
creating an intermediate resonance).

With increasing energy, furthermore, multiparticle production
becomes more and more important.
The high energy collisions -- above $\sqrt{s}$ = 2.6~GeV for
baryon-baryon collisions and $\sqrt{s}$ = 2.2~GeV for meson-baryon
collisions -- are described by  the LUND string fragmentation model
FRITIOF \cite{FRITIOF}. This aspect is similar to that used in the HSD
(Hadron-String-Dynamics) approach \cite{Brat97,CBRep98,Ehehalt} and the UrQMD model
\cite{Bass}.

This combined resonance-string approach allows to calculate
particle production in baryon-baryon and meson-baryon
collisions from low to high energies.
The collisional dynamics for proton-nucleus reactions, furthermore, is
described by the coupled-channel BUU transport approach
\cite{Effe99gam,EffePhD} that is based on the same elementary cross
sections.

\subsection{Dilepton production}

The dilepton production within the resonance model can be schematically
presented in the following way:
\begin{eqnarray}
 BB &\to&R X   \label{chBBR} \\
 mB &\to&R X \label{chmBR} \\
      && R \to  e^+e^- X, \label{chRd} \\
      && R \to  m X, \ m\to e^+e^- X, \label{chRMd} \\
      && R \to  R^\prime X, \ R^\prime \to e^+e^- X, \label{chRprd}
\end{eqnarray}
i.e. in a first step a resonance $R$ might be produced
in baryon-baryon ($BB$) or meson-baryon ($mB$) collisions -- (\ref{chBBR}),
(\ref{chmBR}). Then this resonance can couple to  dileptons
directly -- (\ref{chRd}) (e.g., Dalitz decay of the $\Delta$ resonance:
$\Delta \to e^+e^-N$) or decays to a meson $m$ (+ baryon) -- (\ref{chRMd})
which produces dileptons via direct decays ($\rho, \omega$)
or Dalitz decays ($\pi^0, \eta, \omega$).
The resonance $R$ might also decay into another resonance $R^\prime$ --
(\ref{chRprd}) which later produces dileptons via Dalitz decay or again
via meson decays  (e.g., $D_{35}(1930)\to\Delta\rho,\ \Delta\to
e^+e^-N, \ \rho\to e^+e^-$).  Note, that in the combined model the
final particles -- which couple to dileptons -- can be produced also
via non-resonant mechanisms, i.e. 'background' at low and intermediate
energies and string decay at high energies.

The electromagnetic part of all conventional dilepton sources  --
$\pi^0, \eta, \omega$ and $\Delta$ Dalitz decay, direct decay of vector
mesons $\rho, \omega$ and $pn$ bremsstrahlung -- are treated in the
same way as described in detail in Ref.~\cite{BCM00SIS}-- where dilepton
production in $pp$ and $pd$ reactions has been studied -- and
should not be repeated here again. A  description of the elementary
dilepton sources can be found also in Ref. \cite{Tuebingen}.

\section{In-medium effects on dilepton production.}

\subsection{Collisional broadening and in-medium propagation}

In line with Refs. \cite{GKC97} the effects of collisional
broadening for the vector meson width have been implemented:
\begin{eqnarray}
\Gamma^*_V(M,|\vec p|,\rho)=\Gamma_V(M)
+ \Gamma_{coll}(M,|\vec p|,\rho) ,
\label{gammas}
\end{eqnarray}
where the collisional width is given as
\begin{eqnarray}
\Gamma_{coll}(M,|\vec p|,\rho) = \gamma \ \rho < v \ \sigma_{VN}^{tot} >.
\label{dgamma}
\end{eqnarray}
Here $v=|{\vec p}|/E, \ {\vec p}, \ E$ are the vector
meson velocity, 3-momentum and energy with respect to the target at
rest, $\gamma$ is the Lorentz factor for the boost to the rest frame
of the vector meson, $\rho$ the
nuclear density and $\sigma_{VN}^{tot}$ is the meson-nucleon total
cross section calculated within the Manley resonance model
\cite{Manley}, while $\Gamma_V(M)$ denotes the vacuum width according
to the Manley parametrization \cite{Manley} (for details see
Ref. \cite{Effe99gam}). In Eq.~(\ref{dgamma}) the brackets stand
for an average over the Fermi distribution of the nucleons.

While propagating through the nuclear medium the total width of the
vector meson $\Gamma_V^*$ (\ref{gammas}) changes dynamical and its
spectral function is modified according to the real part of the vector
meson self energy $Re \Sigma^{ret}$, as well as by the imaginary part
of the self energy ($\Gamma_V^*\simeq -Im \Sigma^{ret}/M$) following
\begin{eqnarray}
A_V(M) =  {2\over \pi} \ {M^2 \Gamma_V^*
\over (M^2-M_0^2- Re \Sigma^{ret})^2 + (M {\Gamma_V^*})^2},
\label{spfun}\end{eqnarray}
which is the in-medium form for a boson spectral function.

Since the vector mesons are produced at finite density in line with the
mass-distribution (\ref{spfun}) with $\Gamma_V^* \neq \Gamma_V$ in the
kinematical allowed mass regime, their spectral function has to merge
the vacuum spectral function when propagating out of the medium. To
specify the actual (and general) problem let us consider the decay of
the $N(1520)$ to a nucleon and a $\rho$-meson:  Only a low mass 'slice'
of $\rho$ mesons can be  populated due to energy conservation; such low
mass $\rho$ test-particles can only change their actual mass by
collisions with other hadrons in the approach of \cite{Effe99gam}. In
practice, however, such test-particles do not scatter often enough to
reconstruct the vacuum spectral function when propagating out of the
nucleus. As a consequence such low mass $\rho$'s may propagate to the
vacuum without collisions and radiate dileptons for a long time in the
vacuum since their lifetime, given by the inverse 2 pion decay width,
is very long in the vacuum for low invariant mass.  This 'artefact' is
enhanced by the dilepton radiation probability which -- due to the
virtual photon propagator and a phase-space factor --  is $\sim
M^{-3}$. As a consequence the mass differential dilepton spectrum shows
a large peak close to $M = 2 m_\pi$ (cf. Fig. 14 of \cite{Effe99gam}.
Some 'prescriptions' have been used in \cite{Effe99gam} to cure the
problem: either an 'instantaneous' $\rho$ meson decay, a minimum 2 pion
width of 10 MeV or a mass- and density-dependent real 'potential' for
the $\rho$'s (by Monte Carlo) which reconstructs  the vacuum $\rho$
spectral function when the test-particles propagate out of the nucleus.
The differences between these 'prescriptions' are dramatic for $M \leq
0.4$ GeV but become small for $M \geq$ 0.5 GeV (cf. Fig. 14 of
\cite{Effe99gam}).

In order to avoid the 'low-mass ambiguities' from such 'numerical
prescriptions', which do not appear in perturbative calculational
schemes,  in this study the general off-shell equations of motion from
Refs.  \cite{Cass_off1,Cass_off2} have been employed. Related equations
for the nonrelativistic case have been given in Ref. \cite{Leupold_off}.
In \cite{Cass_off1,Cass_off2} the equations of
motion for test particles  with momentum $\vec P_i$, energy
$\varepsilon_i$ at position $\vec X_i$ -- representing a short-lived
off-shell particle -- have been extended to
\begin{eqnarray}
\label{eomr} \frac{d {\vec X}_i}{dt} \! & = &
\frac{1}{2 \varepsilon_i} \: \left[ \, 2 \,
{\vec P}_i \, + \, {\vec \nabla}_{P_i} \, Re \Sigma^{ret}_{(i)} \,
+ \, \frac{ \varepsilon_i^2 - {\vec P}_i^2 - M_0^2 - Re
\Sigma^{ret}_{(i)}}{\Gamma_{(i)}} \: {\vec \nabla}_{P_i} \,
\Gamma_{(i)} \: \right],
\\[0.3cm]
\label{eomp} \frac{d {\vec P}_i}{d t} \! & = &
\frac{1}{2 \varepsilon_{i}} \: \left[ {\vec
\nabla}_{X_i} \, Re \Sigma^{ret}_i \: + \: \frac{\varepsilon_i^2 -
{\vec P}_i^2 - M_0^{2} - Re \Sigma^{ret}_{(i)}}{\Gamma_{(i)}} \:
{\vec \nabla}_{X_i} \, \Gamma_{(i)} \: \right],
\\[0.3cm]
\label{eome} \frac{d \varepsilon_i}{d t}
& = & \frac{1}{2 \varepsilon_i} \: \left[ \frac{\partial Re
\Sigma^{ret}_{(i)}}{\partial t} \: + \: \frac{\varepsilon_i^2 - {\vec
P}_i^2 - M_0^{2} - Re \Sigma^{ret}_{(i)}}{\Gamma_{(i)}} \:
\frac{\partial \Gamma_{(i)}}{\partial t} \right],
\end{eqnarray}
where the notation $F_{(i)}$ implies that the function is taken at
the coordinates of the test particle, i.e. $F_{(i)} \equiv
F(t,\vec{X}_{i}(t),\vec{P}_{i}(t),\varepsilon_{i}(t))$.
In Eqs. (\ref{eomr})-(\ref{eome}) $Re \Sigma^{ret}$ denotes the real
part of the retarded self energy while $\Gamma = -1/2 Im \Sigma^{ret}$
stands for the imaginary part in short-hand notation. Note, that in
(\ref{eomr})-(\ref{eome}) energy derivatives of the self energy
$\Sigma^{ret}$ have been discarded (cf. \cite{Cass_off1,Cass_off2}).
This should work out well according to the model studies in \cite{Cass_off2}
for the proton-nucleus case.

Furthermore, following Ref. \cite{Cass_off1} and using $M^{2} = P^2 -
Re \Sigma^{ret}$ as an independent variable instead of the energy
$P_0 \equiv \varepsilon$, Eq. (\ref{eome}) turns to
\begin{eqnarray}
\frac{dM_i^2}{dt} \; = \; \frac{M_i^2 -
M_0^2}{\Gamma_{(i)}} \; \frac{d \Gamma_{(i)}}{dt}
\label{eomm}
\end{eqnarray}
for the time evolution of the test-particle $i$ in the invariant
mass squared \cite{Cass_off1,Cass_off2}.

Apart from the propagation in the real potential
$\sim Re \Sigma/2\varepsilon$ the equations (\ref{eomr}) -- (\ref{eomm})
include the dynamical changes due to the imaginary part of the
self energy $Im \Sigma^{ret} \sim - M \Gamma_{V}^*$ with
$\Gamma_V^*$ from (\ref{gammas}).  It is worth to mention that the
deviation from the pole mass, i.e. $\Delta M^2 = M^2 - M_0^2$, follows
the equation
\begin{eqnarray}
{d\over dt}\Delta M^2 = {\Delta M^2\over Im \Sigma^{ret}}
\ {d\over dt} Im\Sigma^{ret},
\label{dm2}\end{eqnarray}
which expresses the fact that the off-shellness in mass is proportional
to the total width $\Gamma_V^*$.  Note, furthermore, that the
equations of motion (\ref{eomr}) -- (\ref{eomm}) conserve the particle
energy $\varepsilon$ if the self energy $\Sigma^{ret}$ does not depend
on time explicitly (cf. Refs.  \cite{Cass_off1,Cass_off2}), which is
approximately the case for $p+A$ reactions.

In this study the effects of collisional broadening described by
$\Gamma_{coll}$ (\ref{dgamma}) with and without an explicit potential
($Re \Sigma^{ret}$) for the vector mesons will be considered (cf. next
Subsection).

\subsection{'Dropping' vector meson mass}

In order to explore the observable consequences of vector meson mass
shifts at finite nuclear density the in-medium vector meson masses are
modeled according to the Hatsuda and Lee \cite{H&L92} or Brown/Rho
scaling \cite{BrownRho} as
\begin{eqnarray}
\label{Brown}
M^* = M_0 \left(1 - \alpha {\rho ({\vec r}) \over \rho_0}\right),
\end{eqnarray}
where $\rho ({\vec r})$ is the nuclear density at the resonance decay,
$\rho_0 = 0.16 \ {\rm fm}^{-3}$ and $\alpha \simeq 0.18$ for the $\rho$ and
$\omega$. The choice (\ref{Brown}) corresponds to
\begin{equation}
Re \Sigma^{ret} = M_0^2 \left( \left(\alpha \frac{\rho}{\rho_0}\right)^2
- 2 \alpha \frac{\rho}{\rho_0}\right)
\end{equation}
in (\ref{eomr}) -- (\ref{eomm}),
which is dominated by the attractive linear term in $\rho/\rho_0$
at nuclear matter density $\rho_0$.

The in-medium vector meson masses $M^*$ (\ref{Brown}) in principle have
to be taken into account in the production part as well as for
absorption reactions and for propagation. This is implemented for the
low energy reactions with nucleon resonances. Note, however, that the
vector mesons produced by the FRITIOF model -- as implemented in the
transport approach \cite{Effe99gam} -- have masses according to the
free spectral function. This approximation might not be severe since
the vector mesons from string decay at high energy have high momenta
with respect to the target nucleus where pole-mass shifts are expected
to be small \cite{Peters,Kondr_rho}.  Furthermore, the $N\rho$-width of
the baryonic resonances at finite density \cite{Effe99gam} has not been
modified.  Such modifications are out of the scope of the present
model.

\section{Numerical results}

\subsection{spatial distribution of $e^+e^-$ production from
vector meson decays}

\begin{figure}[h]
\centerline{\psfig{figure=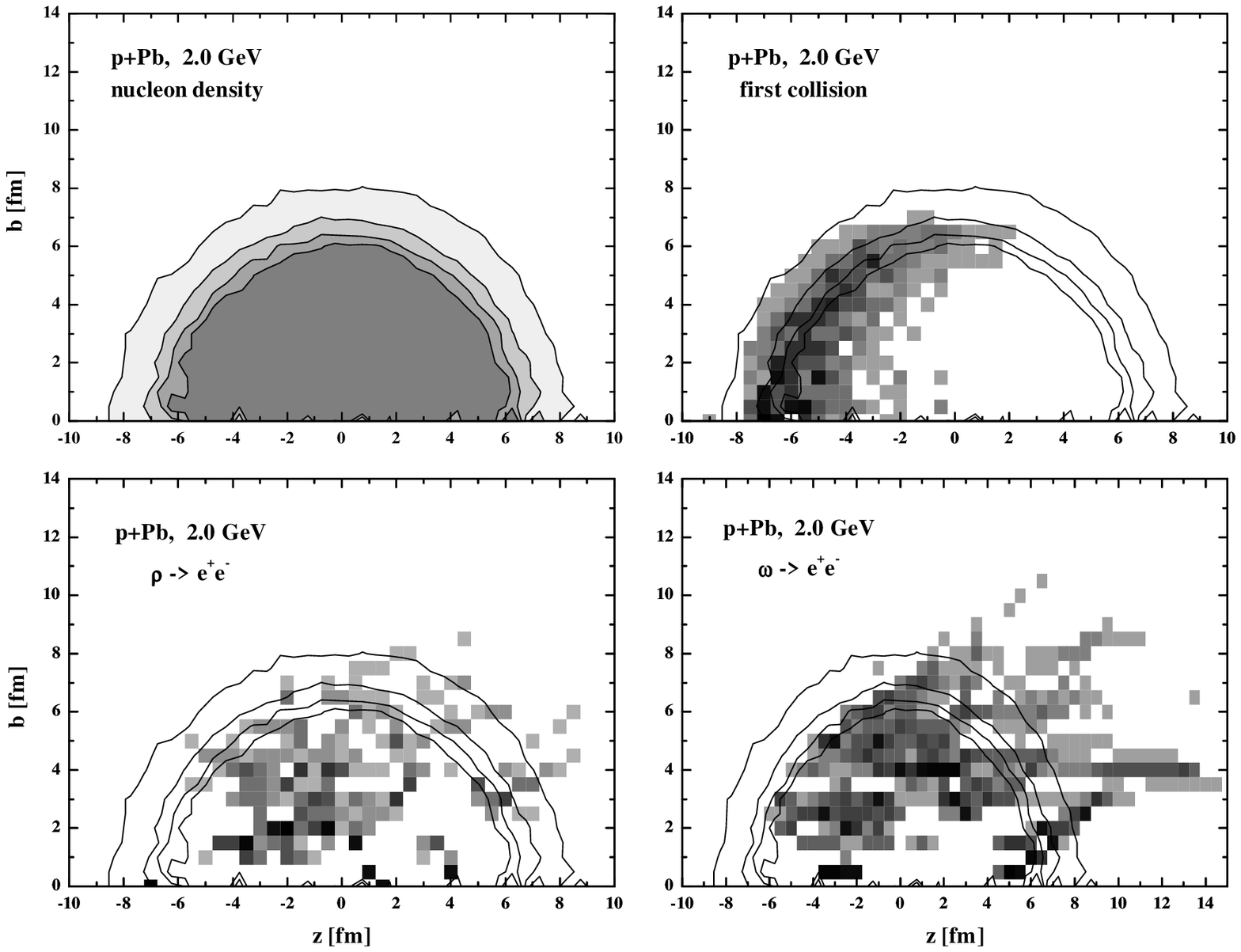,width=15cm}}
\vspace*{-4cm}
\caption{
Upper left part -- the average density distribution of a $Pb$-nucleus at rest
in the laboratory; upper right part -- the spatial distribution
in the first $pN$ collisions; lower part -- the spatial distribution
for $\rho$-meson (left part) and $\omega$-meson (right part) decays
to dileptons. The contour lines correspond to densities of 0.1$\rho_0$,
0.4$\rho_0$, 0.6$\rho_0$ and 0.8$\rho_0$
respectively, and the dark shaded area to $\rho \ge 0.8\rho_0$.}
\label{Fig1pA}
\end{figure}

As mentioned in the introduction dilepton studies from $p+A$ reactions
allow to investigate the vector meson properties at moderate densities
under well controlled conditions and provide complementary information
to $\pi$ or $\gamma$ induced reactions on nuclei. To this aim the
average density distribution of a $Pb$-nucleus at rest in the
laboratory is shown in Fig. \ref{Fig1pA} (upper left part) as well as
the spatial distribution in the first $pN$ collisions (upper right
part).  Here the spatial distribution ${1\over b}{dN\over db dz}$ is
displayed in cylindrical coordinates $b=(x^2+y^2)^{1/2}$ and $z$, where
$z$ is directed along the beam axis and the proton is impinging on the
nucleus from the left side.

The lower part of Fig. \ref{Fig1pA} displays the spatial
distribution for $\rho$-meson (left part) and $\omega$-meson (right
part) decays to dileptons. At low bombarding energy most of the
$\rho$'s stem from the decay of baryonic resonances formed in primary
$pN$ and secondary $\pi N$ collisions. Thus, $\rho$'s are produced
inside the nucleus close to the surface, i.e. at normal nuclear density.
Since $\rho$-mesons have a short life time, they decay to dileptons
basically inside the nucleus.  On the contrary, the $\omega$-mesons are
formed dominantly in primary $pN$ collisions and have a longer life
time, such that the $\omega$ spatial distribution is more elongated and
part of the $\omega$'s decay  outside the nucleus.

\subsection{$p + A$ collisions from 1--4 GeV}

\subsubsection{Invariant mass distributions}

\begin{figure}[t]
\phantom{a}\vspace*{-5mm}
\centerline{\psfig{figure=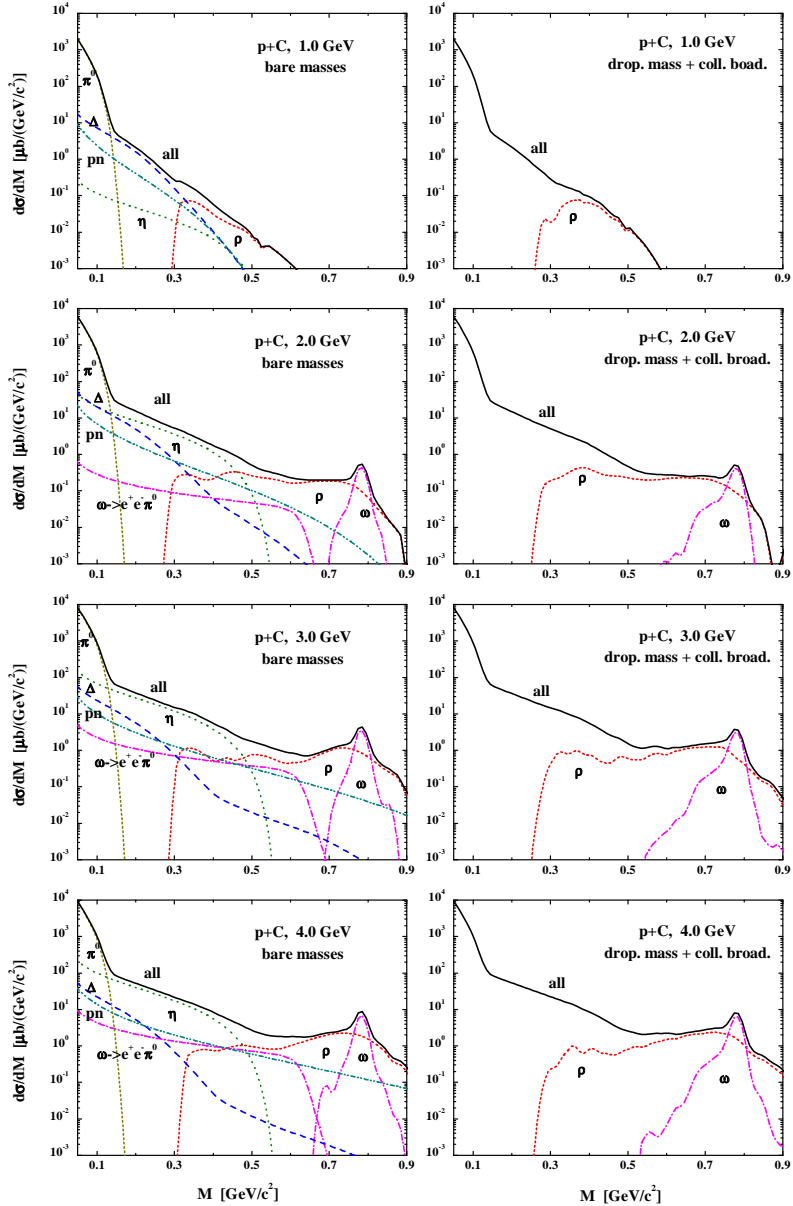,width=12.5cm}}
\vspace*{-0.5cm}
\caption{
The calculated dilepton invariant mass spectra $d\sigma/dM$ for $p + C$
collisions from 1 -- 4 GeV (including an experimental mass
resolution of 10 MeV) without in-medium modifications (bare masses) --
left part, and applying the collisional broadening + dropping mass
scenario -- right part.
The thin lines indicate the individual contributions from the different
production channels; {\it i.e.}~ starting from low $M$:
Dalitz decay $\pi^0 \to \gamma e^+ e^-$ (short dashed line),
$\eta \to \gamma e^+ e^-$ (dotted line),
$\Delta \to N e^+ e^-$ (dashed line),
$\omega \to \pi^0 e^+ e^-$ (dot-dashed line),
for $M \approx $ 0.7 GeV: $\omega \to e^+e^-$ (dot-dashed line),
$\rho^0 \to e^+e^-$ (short dashed line).
The full solid line represents the sum of all sources considered here.}
\label{Fig2pA}e
\end{figure}

In Fig.~\ref{Fig2pA} the calculated dilepton invariant mass spectra
$d\sigma/dM$ are presented for $p + C$ collisions from
1.0 -- 4 GeV (including an experimental mass resolution
$\Delta M$ = 10 MeV) without in-medium modifications (bare masses) --
left part, and applying the collisional broadening + dropping mass
scenario -- right part.
The thin lines indicate the individual contributions from the different
production channels; {\it i.e.}~ starting from low $M$:
Dalitz decay $\pi^0 \to \gamma e^+ e^-$ (short dashed line),
$\eta \to \gamma e^+ e^-$ (dotted line),
$\Delta \to N e^+ e^-$ (dashed line),
$\omega \to \pi^0 e^+ e^-$ (dot-dashed line),
for $M \approx $ 0.7 GeV: $\omega \to e^+e^-$ (dot-dashed line),
$\rho^0 \to e^+e^-$ (short dashed line).
The full solid line represents the sum of all sources considered here.
The dominant contribution at low $M$ ($> m_{\pi^0}$) is
the $\eta$ Dalitz decay, however, for $M > 0.4$ GeV the dileptons stem
basically all from direct vector meson decays ($\rho$ and $\omega$).
Note, that for the collisional broadening + dropping mass scenario
(right panel) only the $\rho$ and $\omega$ contributions as well as the
sum of all sources are presented since the other individual
contributions are similar to the bare mass case (left panel).

It is worth to point out, that already the free $\rho$-contribution is
very asymmetric in mass due to the fact that the dilepton decay leads
to a multiplication of the $\rho$-spectral function by $1/M^3$ (cf.
Ref.  \cite{BCM00SIS}).  This is in contrast to the assumption made in
Ref.  \cite{Japen} for the $e^+e^-$ spectra from $p + C$ and $p + Cu$
reactions at 12 GeV/c laboratory momentum.

\begin{figure}[t]
\phantom{a}\vspace*{5mm}
\centerline{\psfig{figure=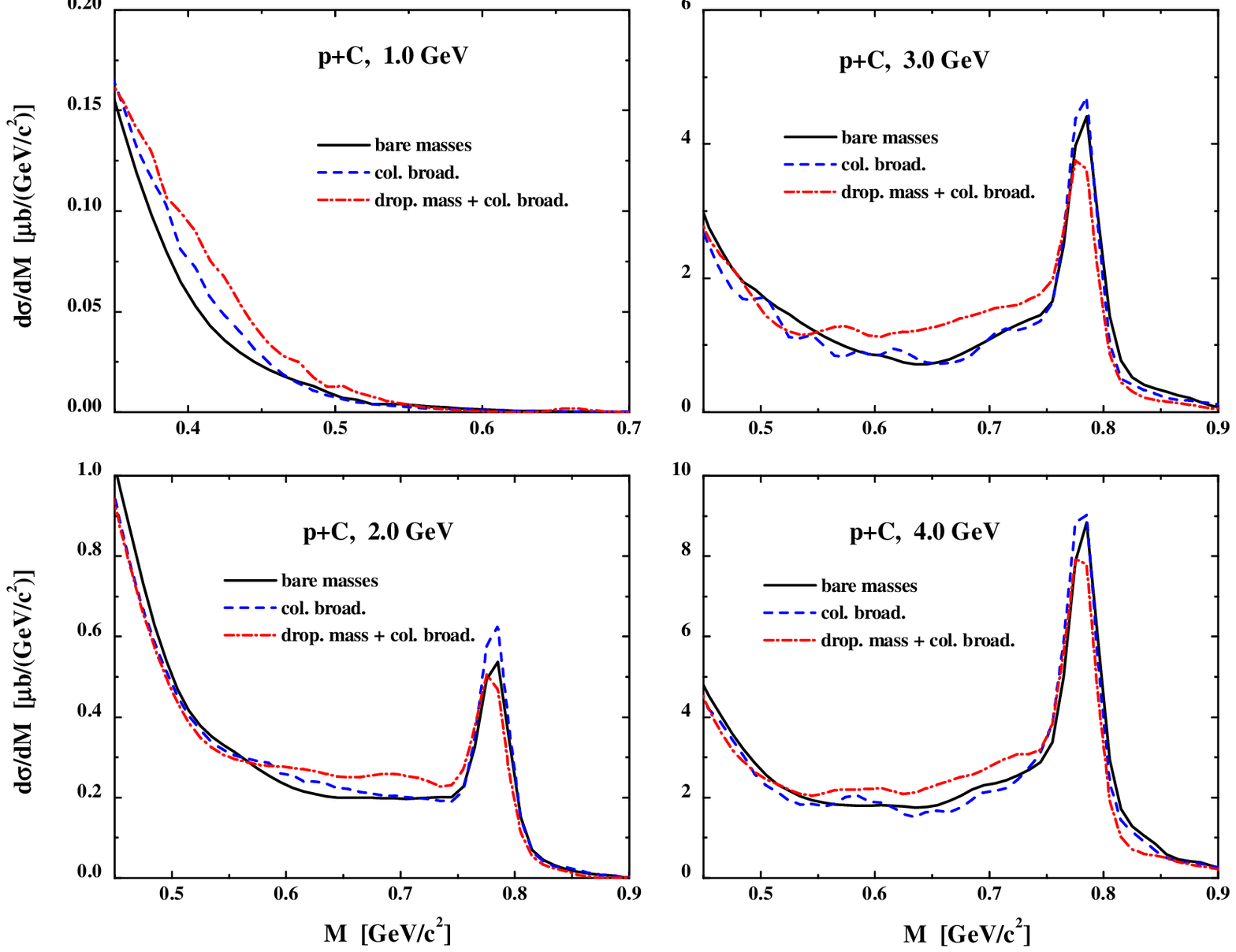,width=15cm}}
\vspace*{-8cm}
\caption{
The comparison of different in-medium modification scenarios, i.e.
collisional broadening (dashed lines) and collisional broadening +
dropping vector meson masses (dash-dotted lines), with respect to the
bare mass case (solid lines) on a linear scale for $p + C$ from 1--4 GeV.}
\label{Fig3pA}
\end{figure}

 In order to see the differences between the results from the left and
right panels of Fig. \ref{Fig2pA},  a comparison of the different
in-medium modification scenarios is shown in Fig. \ref{Fig3pA}, i.e.
collisional broadening (dashed lines) and collisional broadening +
dropping vector meson masses (dash-dotted lines), with respect to the
bare mass case (solid lines) on a linear scale for $p + C$ from 1--4
GeV.  At 1.0 GeV some enhancement for $0.4 \leq M \leq 0.5$ GeV in case
of collisional broadening (dashed line) is found as well as an
additional mass shift (dash-dotted line) which is essentially due to
'subthreshold' $\rho$ production in the $\pi N \rightarrow N^*(1520)
\rightarrow \rho N$ or $pN \rightarrow N N^*(1520) \rightarrow NN \rho$
reactions, where the $\rho$ is stronger populated from the $N^*(1520)$
resonance in case of a broadened (and shifted) $\rho$ spectral
function. The modifications of the dilepton spectrum are rather
moderate for the light $C$-target at 2, 3 and 4 GeV especially for the
collisional broadening scenario since the $\rho$-nucleon-resonance
couplings are already included dynamically in the transport model. Only
in case of the additional $\rho$ mass shift (dash-dotted lines) one
observes a small enhancement for $0.5 \leq M \leq 0.75$ GeV, which is
most pronounced at 3 GeV.

\begin{figure}[t]
\centerline{\psfig{figure=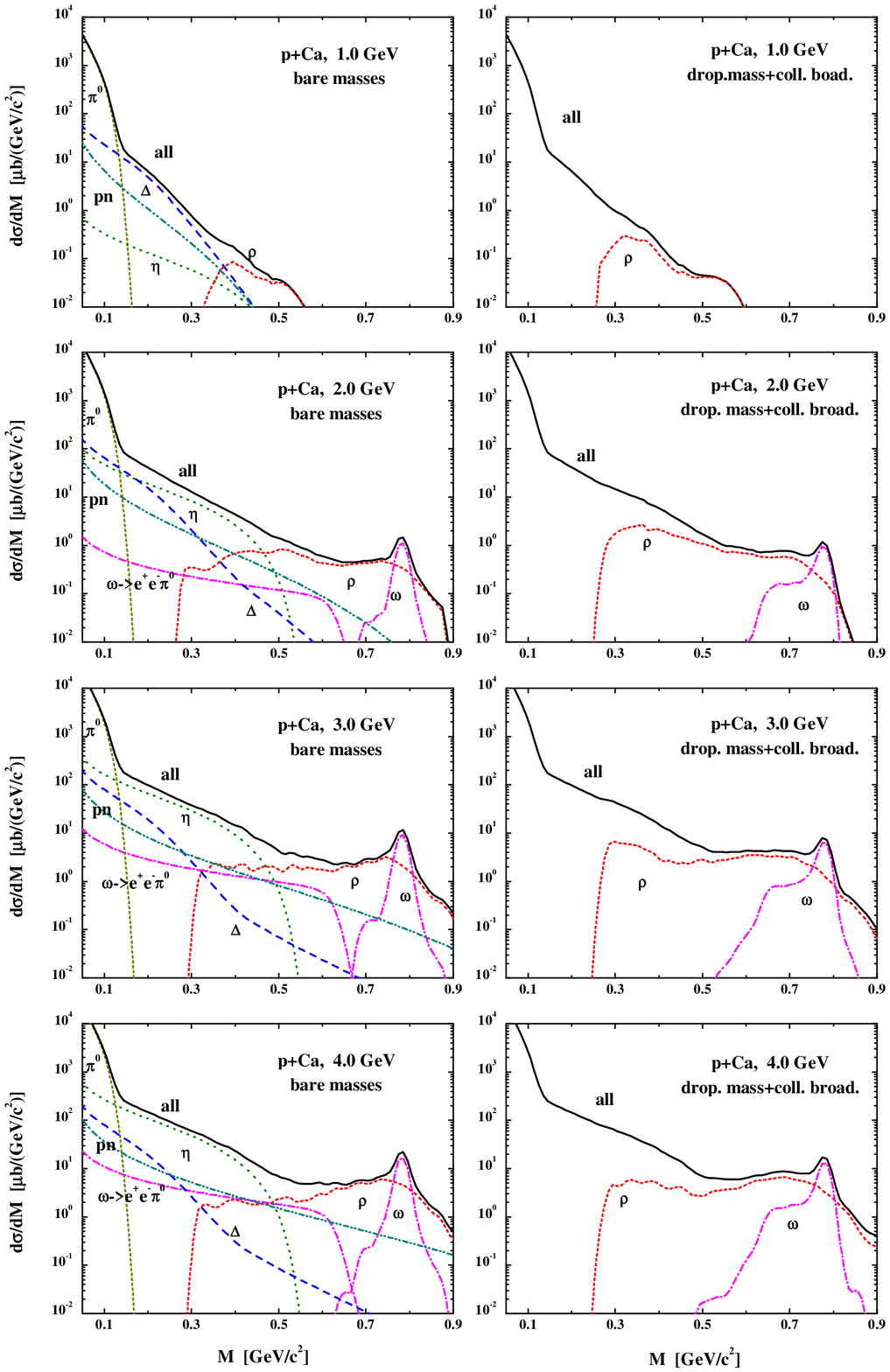,width=12.5cm}}
\vspace*{-5mm}
\caption{
The calculated dilepton invariant mass spectra $d\sigma/dM$ for $p + Ca$
collisions from 1.0 -- 4 GeV (including an experimental mass
resolution of 10 MeV)
without in-medium modifications (bare masses) -- left part,
and applying the collisional broadening + dropping masses scenario
-- right part.
The assignment of the individual lines is the same as in
Fig. \protect\ref{Fig2pA}.}
\label{Fig4pA}
\end{figure}

\begin{figure}[t]
\phantom{a}\vspace*{5mm}
\centerline{\psfig{figure=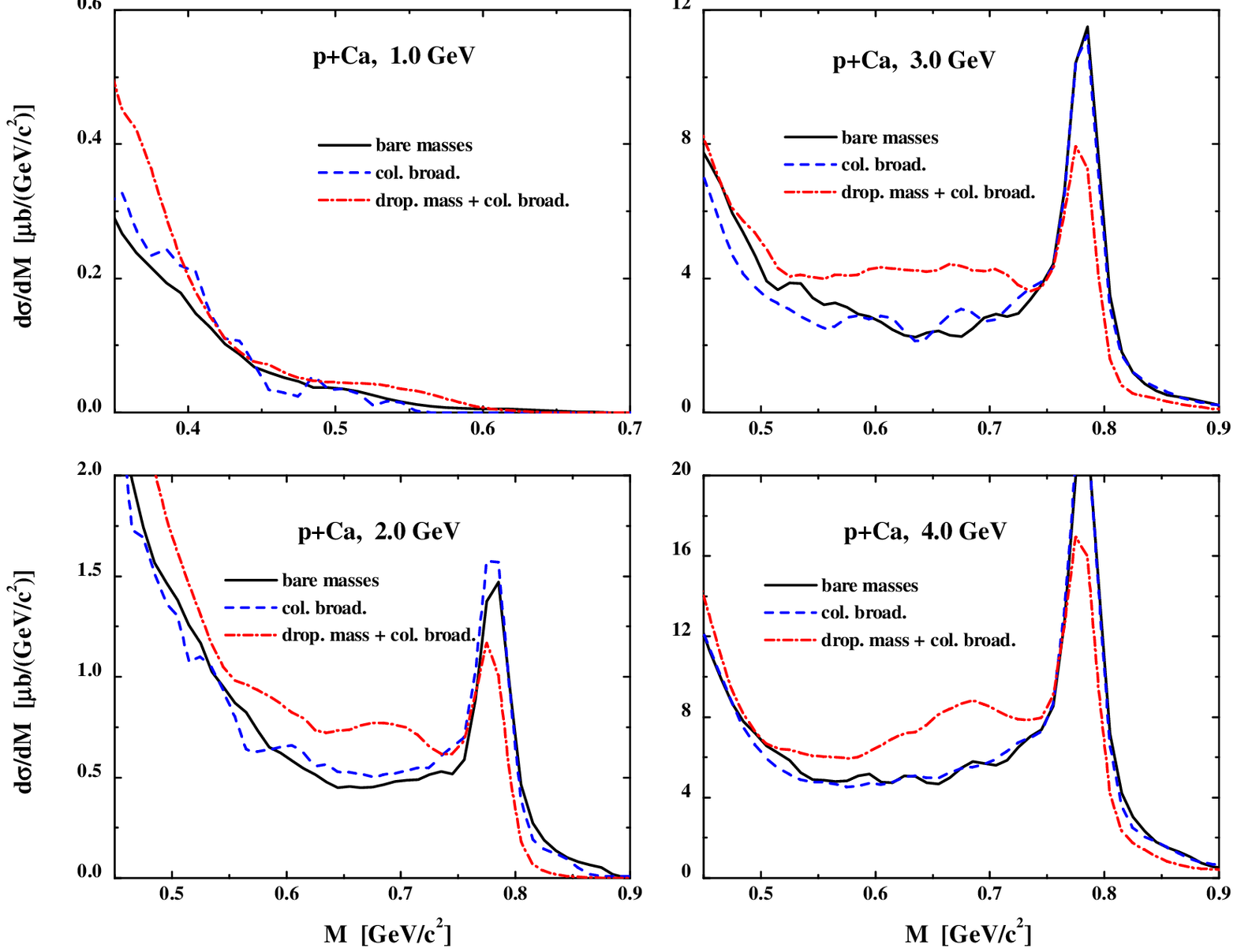,width=15cm}}
\vspace*{-8cm}
\caption{
The comparison of different in-medium modification scenarios, i.e.
collisional broadening (dashed lines) and collisional broadening +
dropping vector meson masses (dash-dotted lines), with respect to the
bare mass case (solid lines) on a linear scale for $p + Ca$ from 1--4 GeV.}
\label{Fig5pA}
\end{figure}

In Figs. \ref{Fig4pA}--\ref{Fig5pA} the calculational results for the
$p + Ca$ reaction from 1--4 GeV are presented in analogy to Figs.
\ref{Fig2pA} and \ref{Fig3pA}. Consequently, the assignment of the
individual lines is the same as in Figs. \ref{Fig2pA}, \ref{Fig3pA} for
the $p + C$ case.  Except from an overall scaling in height, these
spectra look very much the same as for the light $C$ target. The
in-medium modifications again can be much better seen by a direct
comparison on a linear scale in Fig. \ref{Fig5pA}. As mentioned before,
collisional broadening of the $\rho$ spectral function gives no net
signal within the numerical statistics achieved whereas the 'dropping
mass' scenario leads to a now more pronounced enhancement in the range
$0.5 \leq M \leq 0.75$ GeV, which is up to a factor 1.7 at 3 GeV. Note,
that this enhancement correlates with a reduction of the dilepton yield
in the pole-mass region of the $\rho$ and $\omega$ mesons.

Let's continue with the heavy $Pb$ target. In Figs.
\ref{Fig6pA}--\ref{Fig7pA}  the calculated results for $p + Pb$ from
1--4 GeV are presented again in analogy to Figs. \ref{Fig2pA} and
\ref{Fig3pA} with the same assignment of the individual lines.
Apart from an overall scaling in height, these spectra look again very
similar as for the $C$ and $Ca$ targets. The in-medium modifications
again can be much better seen by a direct comparison on a linear scale
in Fig. \ref{Fig7pA}. Now the in-medium effects show up more clearly.
Whereas collisional broadening of the $\rho$ spectral function again
gives no clear signal within the numerical accuracy achieved the
'dropping mass' scenario leads to a pronounced modification of the
spectral shape. A strong reduction of the dilepton yield in the vector
meson pole mass region around 0.77 GeV is observed since most of
the $\rho$'s and $\omega$'s now decay in the medium approximately at
density $\rho_0$. This leads to a pronounced peak around $M \approx
0.65$ GeV, which can be attributed to the in-medium $\omega$ decay
since the $\rho$ spectral strength is distributed over a wide low mass
regime. The situation is very reminiscent of dilepton spectra from $\pi
+ A$ and $\gamma + A$ reactions in Refs.
\cite{CBRep98,Effe_piA,Effe99gam}.  Especially when comparing dilepton
spectra from $C$ and $Pb$ targets, it should be experimentally possible
to distinguish an in-medium mass shift of the $\omega$ meson by taking
the ratio of both spectra.

In summarizing the results from Figs. \ref{Fig3pA}, \ref{Fig5pA} and
\ref{Fig7pA}, the collisional broadening scenario gives practically the
same dilepton spectra  (within statistical fluctuations) as the bare
mass case, since the coupling of the $\rho$'s to the baryonic
resonances are dynamicaly taken into account in the resonance model.
The inclusion of 'dropping' vector meson masses leads to an enhancement
of the dilepton yield for $M=0.5-0.75$ GeV and to a reduction at
the $\omega$-peak which becomes more pronounced with increasing target size
and indicates a factor of 2 enhancement from in-medium $\omega$ decays
in case of $p + Pb$ at 3--4 GeV.

\subsubsection{Transverse momentum distributions}

In Fig. \ref{Fig8pA} the transverse momentum distribution of all
dileptons for the $p + C$ system at 1.0 , 2.0, 3.0 and 4.0 GeV are
displayed. Here the sequence of the individual channels is as follows:
the contribution from the $\pi^0$ Dalitz decay dominates at all
energies and is practically identical to the sum of all sources (solid
line). Apart from 1.0 GeV the next strong channel is the $\eta$ Dalitz
decay followed by the $\Delta$-Dalitz decays, $pn$ bremsstrahlung and
the $\omega$-Dalitz decays, while the direct decays of the $\rho$ and
$\omega$ mesons are down by orders of magnitude. Thus global $p_T$
spectra do not provide very interesting information.

In order to extract the interesting physics one has to apply cuts on
the invariant dilepton mass to suppress or exclude the dominant
channels. For this purpose a cut on the interval $0.4 \leq M \leq 0.7$
GeV has been chosen, which is displayed for the $p + C$ reaction at
2.0, 3.0 and 4.0 GeV in Fig. \ref{Fig9pA} (left panel) for the bare
mass case. Indeed, now the $\rho$ meson signal is the strongest or at
least comparable to the remaining $\eta$ Dalitz decay, whereas the
other channels are more suppressed. In the right panel of Fig.
\ref{Fig9pA} a comparison of the bare mass case (solid lines) with the
collisional broadening and dropping mass scenario (dash-dotted lines)
is presented.  Since the differences are only very tiny, one can
conclude that there is practically no sensitivity to in-medium effects
for the $C$ target at all energies.

Following the same strategy as in the previous subsection  the
same analysis for the system $p + Ca$ is shown in Figs. \ref{Fig10pA} and
\ref{Fig11pA}.  The $p_T$ spectrum of the dileptons is again similar at
 all energies to that from the $C$ target in Figs. \ref{Fig8pA} and
\ref{Fig9pA}, however, the detailed comparison of the bare mass (solid
line) and dropping mass + collisional broadening scenario (dash-dotted
line) in Fig.  \ref{Fig11pA}  (right panel) indicates an enhancement of
the $p_T$ spectrum in case of the in-medium modifications whereas the
shape in $p_T$ is very similar.  These observations are practically
identical even for the heavy $Pb$ target as demonstrated in Figs.
\ref{Fig12pA} and \ref{Fig13pA}. Here the $p_T$ spectrum from the
different scenarios (right panel in Fig. \ref{Fig13pA}) show the same
enhancement for the dropping mass case as in Fig. \ref{Fig11pA},
however, the spectral shape in $p_T$ does not provide new information
since it does not differ significantly at any transverse momentum.

\subsubsection{Dilepton rapidity distributions}

In Fig. \ref{Fig14pA}  the rapidity distribution of all dileptons in
the laboratory for the $p + C$ system at 1.0 , 2.0, 3.0 and 4.0 GeV is
displayed.  Here the sequence of the individual lines is follows: the
contribution from the $\pi^0$ Dalitz decay dominates at all energies
and again is almost identical to the sum of all sources (solid
line). Apart from 1.0 GeV the next strong channel is the $\eta$ Dalitz
decay followed by the $\Delta$-Dalitz decays, $pn$ bremsstrahlung and
the $\omega$-Dalitz decays, while the direct decays of the $\rho$ and
$\omega$ mesons are again barely visible.

In order to extract the interesting information cuts on the invariant
dilepton mass are necessary to suppress/exclude the dominant channels.
Again a cut on the interval $0.4 \leq M \leq 0.7$ GeV has been chosen,
which is displayed for the $p + C$ reaction at 2.0, 3.0 and 4.0 GeV in
Fig. \ref{Fig15pA} (left panel) for the bare mass case. Indeed, now the
$\rho$ meson signal is the strongest or comparable to the remaining
$\eta$ Dalitz decay (which will be even more suppressed by gating on
the interval $0.55 \leq M \leq 0.75$ GeV).  In the right panel of Fig.
\ref{Fig15pA} a comparison of the bare mass case (solid lines) with the
collisional broadening and dropping mass scenario (dash-dotted lines)
is presented.  As in case of the $p_T$ spectra the differences are only
very tiny;  one thus can conclude that there is no sensitivity
to in-medium effects for the $C$ target with respect to rapidity
spectra, too.

For completeness, in Figs. \ref{Fig16pA} to \ref{Fig19pA} the same
analysis is shown for the $Ca$ and $Pb$ target at all energies. As
seen from the right panels in Figs. \ref{Fig17pA} and \ref{Fig19pA} the
shape of the rapidity distribution is not changed very much for the
in-medium mass scenario compared to the bare mass case except for a
tiny shift to lower laboratory rapidities. As in case of the $p_T$
spectra in the previous subsection, an overall enhancement in the
dilepton yield for the in-medium mass scenario for the $p + Pb$
reaction is the most pronounced effect.

\subsubsection{Double differential dilepton spectra}

\begin{figure}[t]
\centerline{\psfig{figure=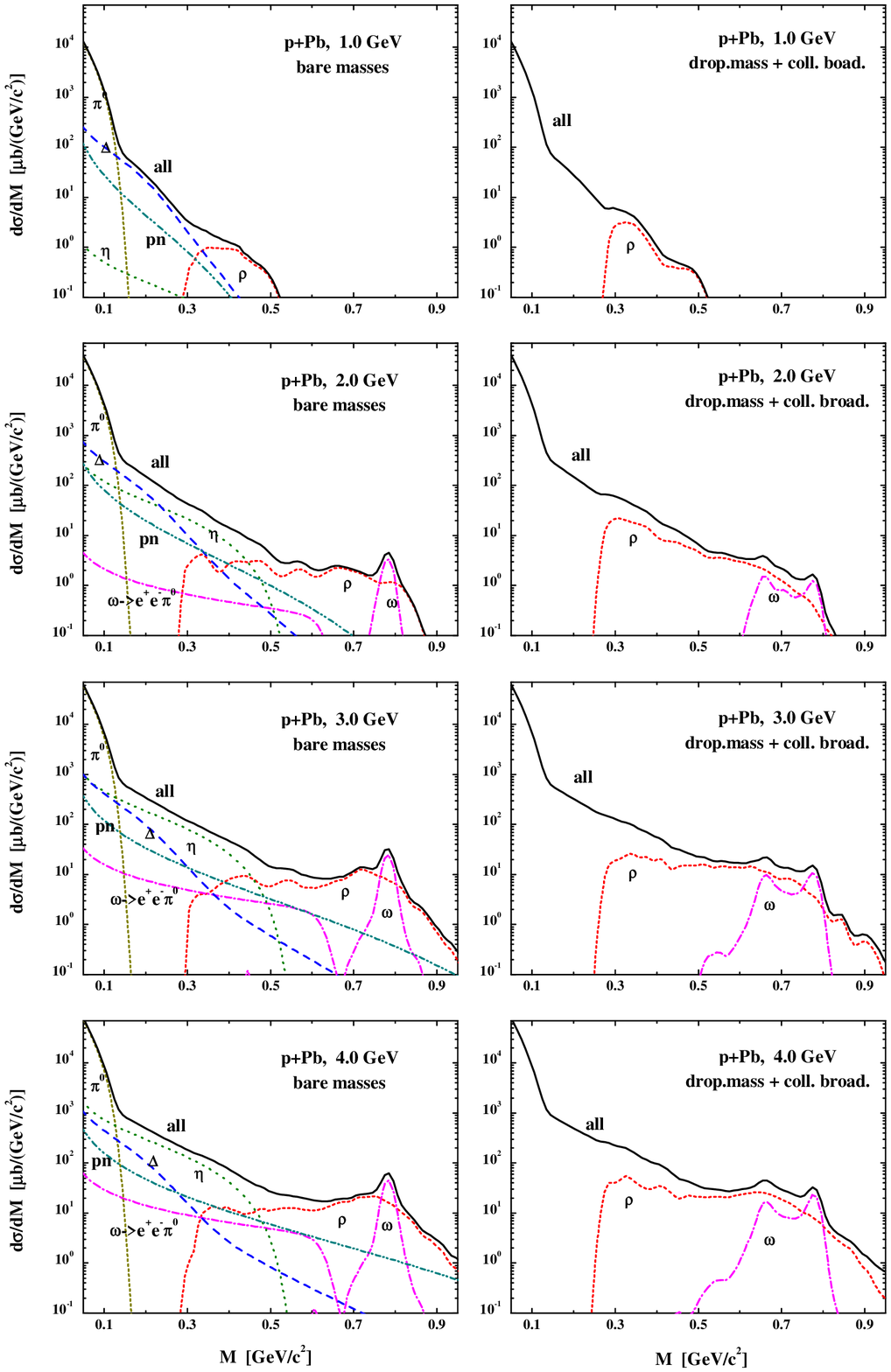,width=12.5cm}}
\vspace*{-5mm}
\caption{
The calculated dilepton invariant mass spectra $d\sigma/dM$ for $p + Pb$
collisions from 1.0 -- 4 GeV (including an experimental mass
resolution of 10 MeV)
without in-medium modifications (bare masses) -- left part,
and applying the collisional broadening + dropping masses scenario
-- right part.
The assignment of the individual lines is the same as in
Fig. \protect\ref{Fig2pA}.}
\label{Fig6pA}
\end{figure}

\begin{figure}[t]
\phantom{a}\vspace*{5mm}
\centerline{\psfig{figure=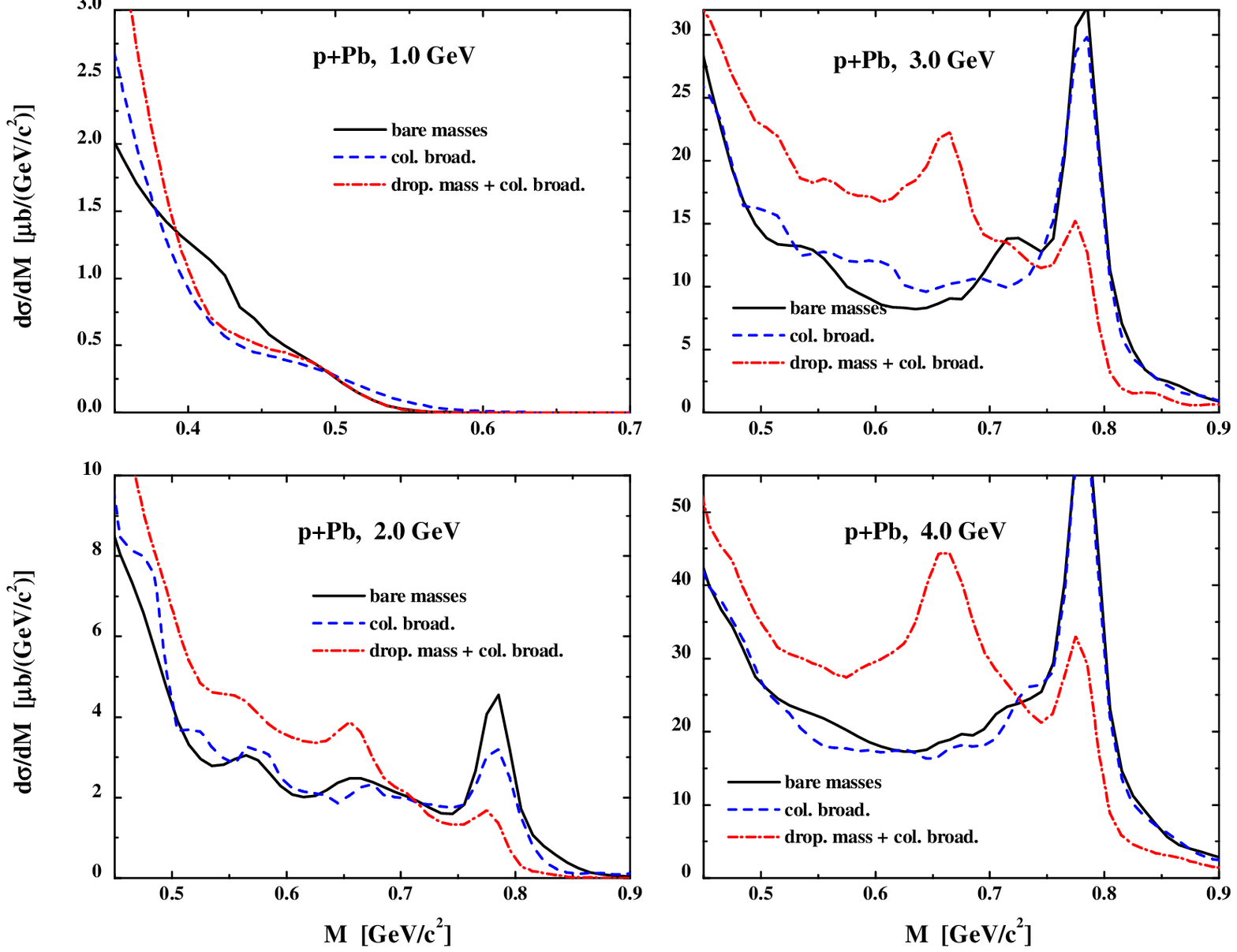,width=15cm}}
\vspace*{-8cm}
\caption{
The comparison of different in-medium modification scenarios, i.e.
collisional broadening (dashed lines) and collisional broadening +
dropping vector meson masses (dash-dotted lines), with respect to the
bare mass case (solid lines) on a linear scale for $p + Pb$ from 1--4 GeV.}
\label{Fig7pA}
\end{figure}

\begin{figure}[t]
\phantom{a}\vspace*{5mm}
\centerline{\psfig{figure=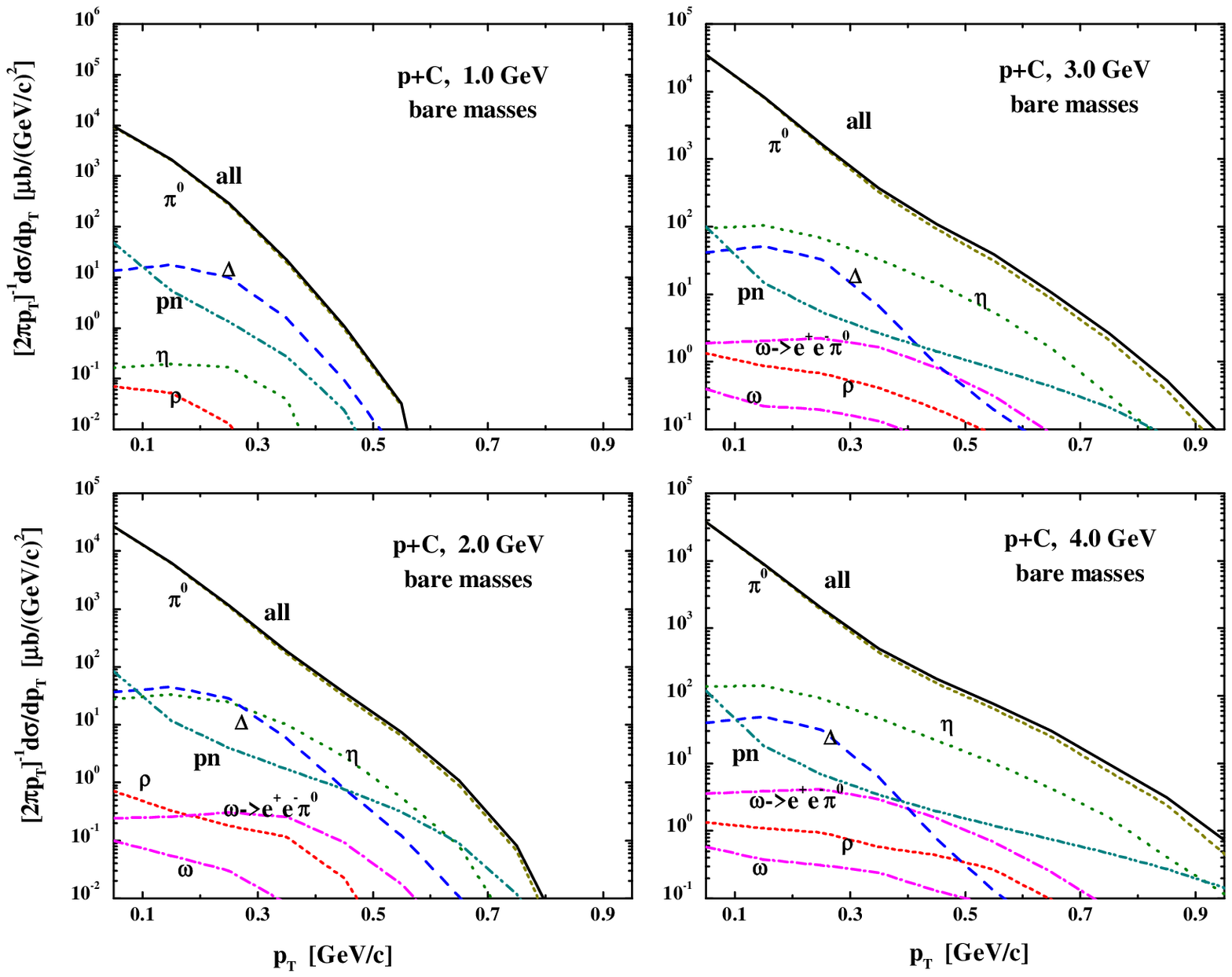,width=15cm}}
\vspace*{-8cm}
\caption{
The transverse momentum distribution $d\sigma/dp_T /(2\pi p_T)$
for the $p + C$ system at 1.0, 2.0, 3.0 and 4.0 GeV.}
\label{Fig8pA}
\end{figure}

\begin{figure}[t]
\phantom{a}\vspace*{-20mm}
\centerline{\psfig{figure=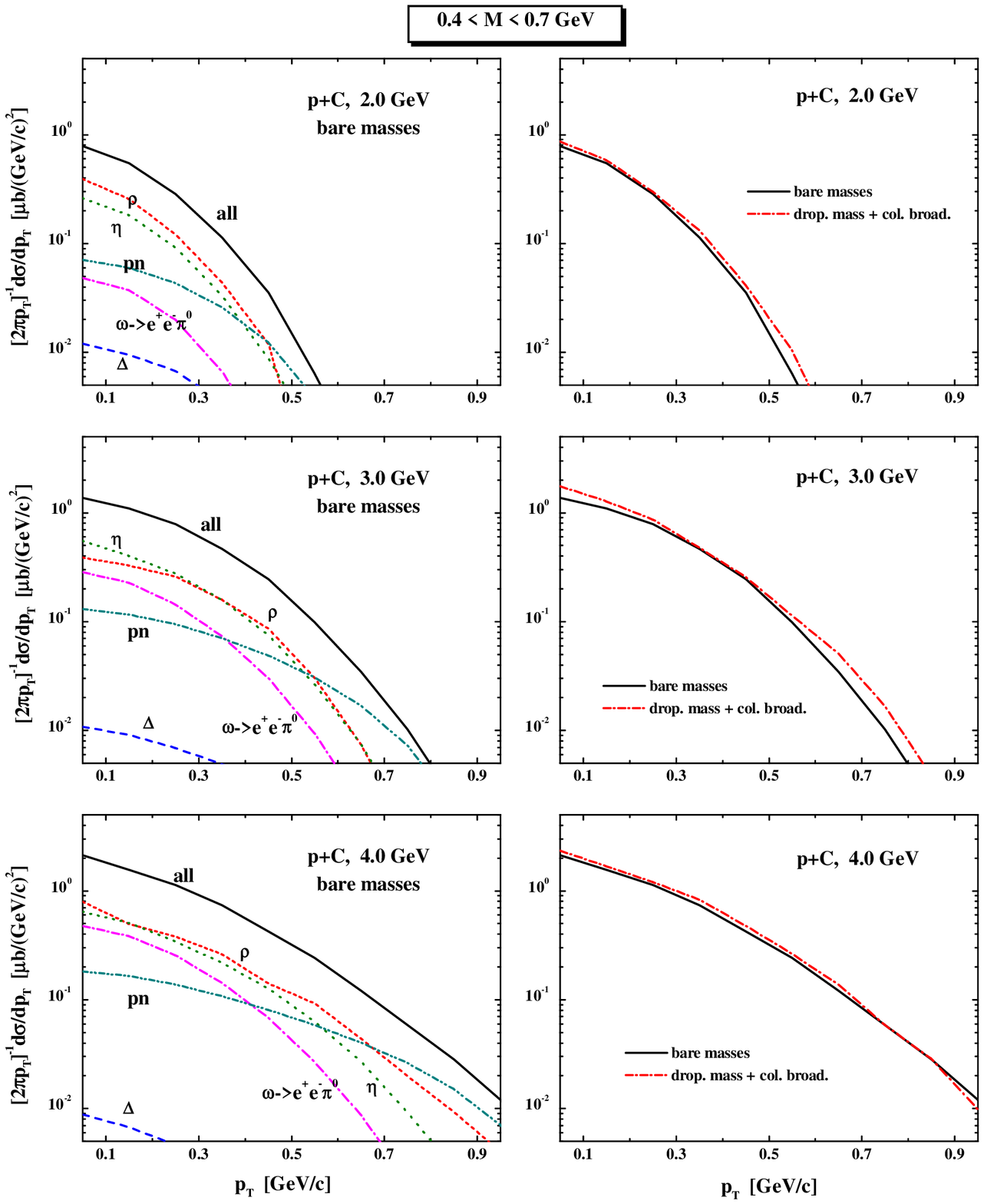,width=15cm}}
\caption{
The transverse momentum distribution $d\sigma/dp_T /(2\pi p_T)$ for the
$p + C$ system at 1.0, 2.0, 3.0 and 4.0 GeV implying a cut in invariant
mass of $0.4 \le M \le 0.7$ GeV.  Left panel -- the individual
contributions for bare mass case, right panel - comparison of the bare
mass spectra (solid line) with the collisional broadening and dropping
mass scenario (dash-dotted lines). }
\label{Fig9pA}
\end{figure}

\begin{figure}[t]
\phantom{a}\vspace*{5mm}
\centerline{\psfig{figure=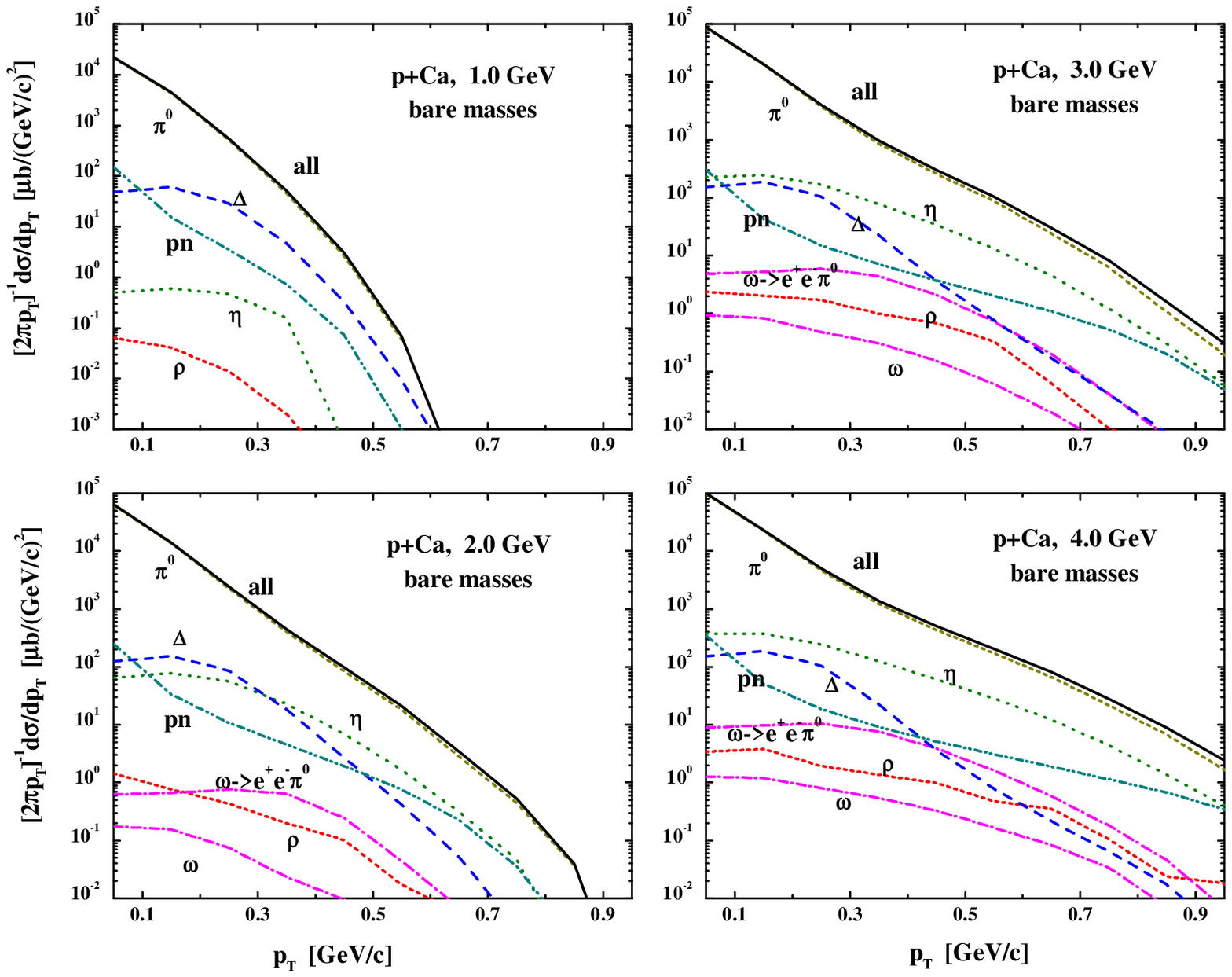,width=15cm}}
\vspace*{-8cm}
\caption{
The transverse momentum distribution $d\sigma/dp_T /(2\pi p_T)$
for the $p + Ca$ system at 1.0, 2.0, 3.0 and 4.0 GeV.}
\label{Fig10pA}
\end{figure}

\begin{figure}[t]
\phantom{a}\vspace*{-20mm}
\centerline{\psfig{figure=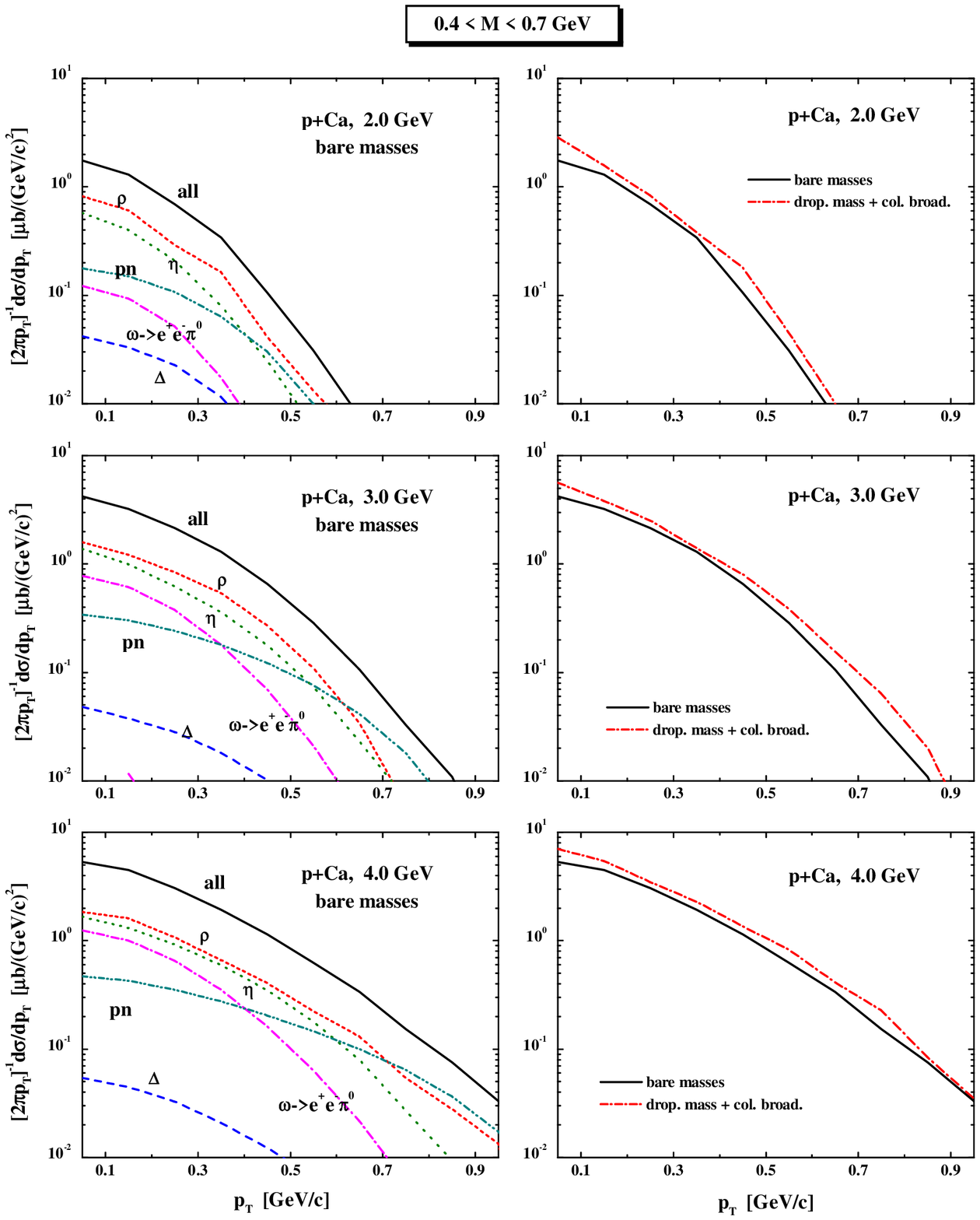,width=15cm}}
\caption{
The transverse momentum distribution $d\sigma/dp_T /(2\pi p_T)$ for the
$p + Ca$ system at 1.0, 2.0, 3.0 and 4.0 GeV implying a cut in invariant
mass of $0.4 \le M \le 0.7$ GeV.  Left panel -- the individual
contributions for bare mass case, right panel - comparison of the bare
mass spectra (solid line) with the collisional broadening and dropping
mass scenario (dash-dotted lines). }
\label{Fig11pA}
\end{figure}

\begin{figure}[t]
\phantom{a}\vspace*{5mm}
\centerline{\psfig{figure=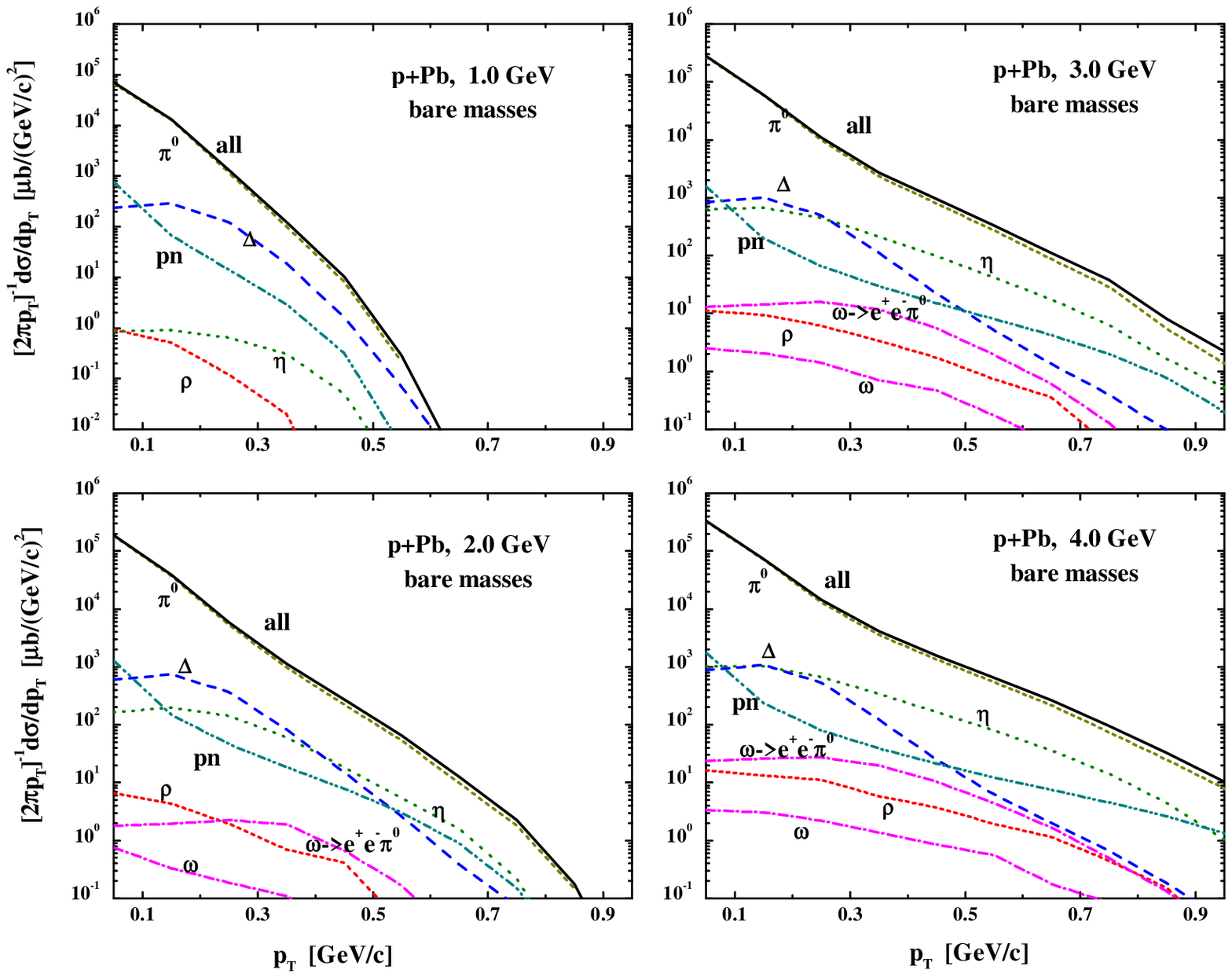,width=15cm}}
\vspace*{-8cm}
\caption{
The transverse momentum distribution $d\sigma/dp_T /(2\pi p_T)$
for the $p + Pb$ system at 1.0, 2.0, 3.0 and 4.0 GeV.}
\label{Fig12pA}
\end{figure}

\begin{figure}[t]
\phantom{a}\vspace*{-20mm}
\centerline{\psfig{figure=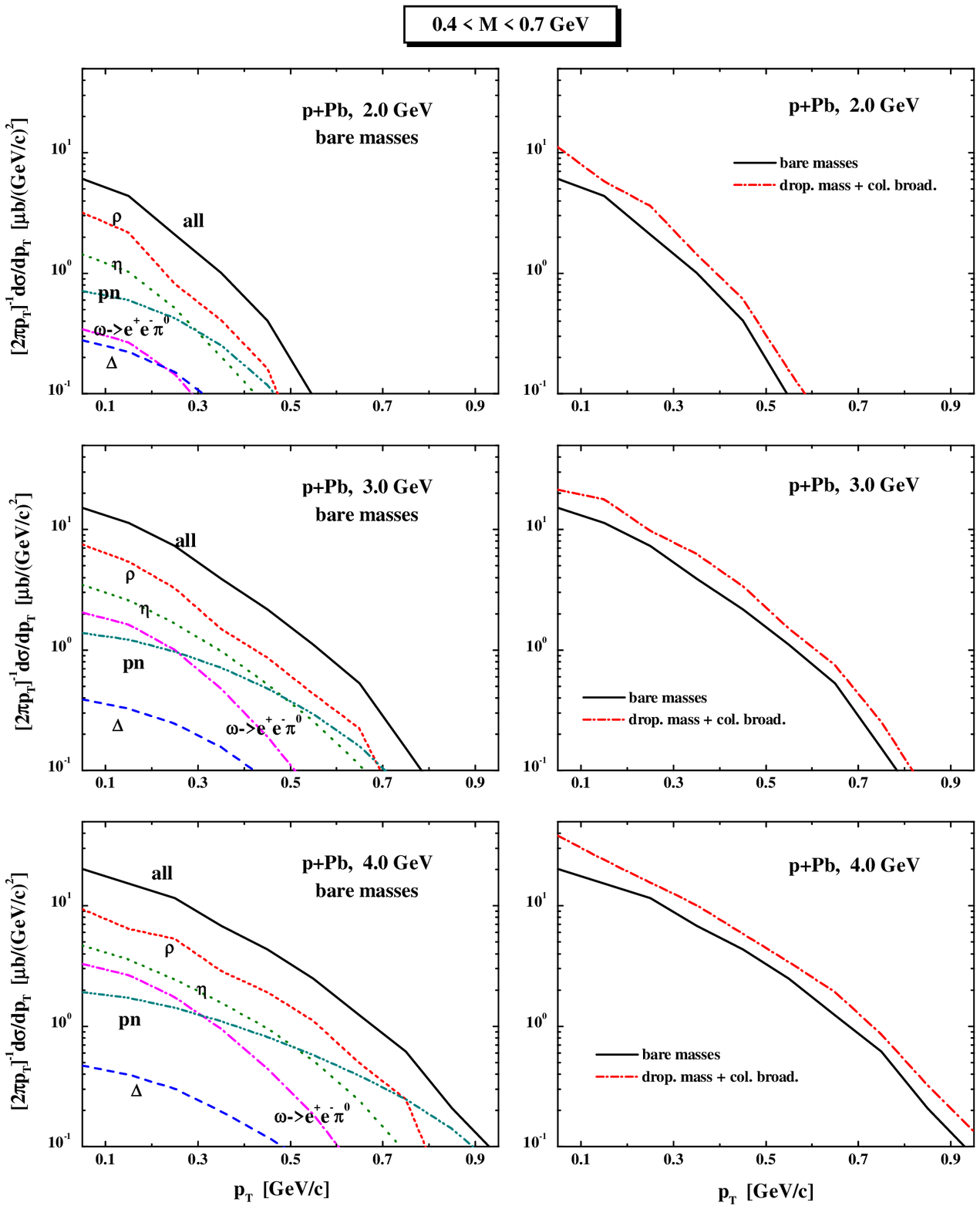,width=15cm}}
\caption{
The transverse momentum distribution $d\sigma/dp_T /(2\pi p_T)$ for the
$p + Pb$ system at 1.0, 2.0, 3.0 and 4.0 GeV implying a cut in invariant
mass of $0.4 \le M \le 0.7$ GeV.  Left panel -- the individual
contributions for bare mass case, right panel - comparison of the bare
mass spectra (solid line) with the collisional broadening and dropping
mass scenario (dash-dotted lines). }
\label{Fig13pA}
\end{figure}

\begin{figure}[t]
\phantom{a}\vspace*{5mm}
\centerline{\psfig{figure=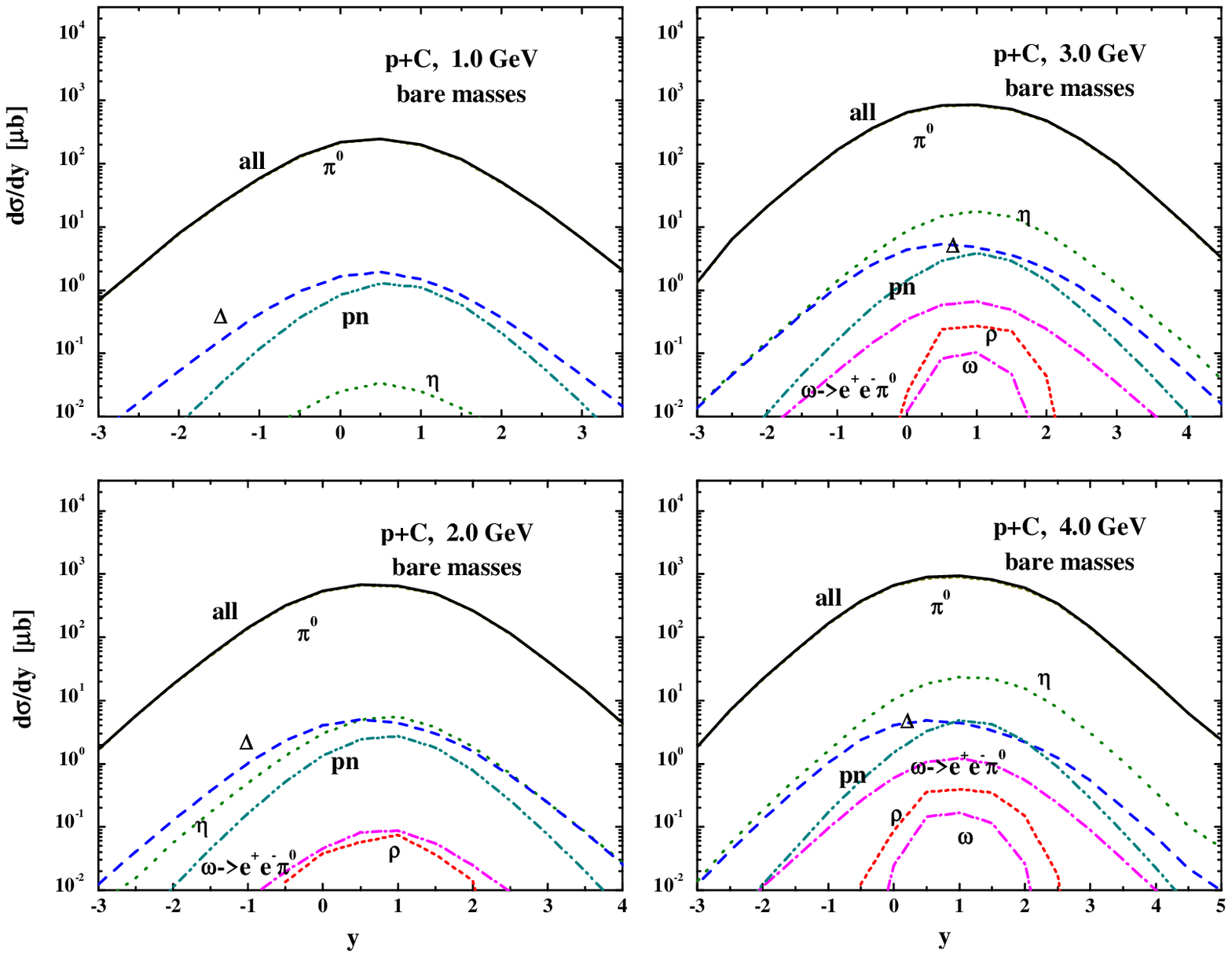,width=15cm}}
\vspace*{-8cm}
\caption{
The laboratory rapidity distributions $d\sigma/dy$ for the $p + C$ system
at 1.0, 2.0, 3.0 and 4.0 GeV. }
\label{Fig14pA}
\end{figure}

\begin{figure}[t]
\phantom{a}\vspace*{-20mm}
\centerline{\psfig{figure=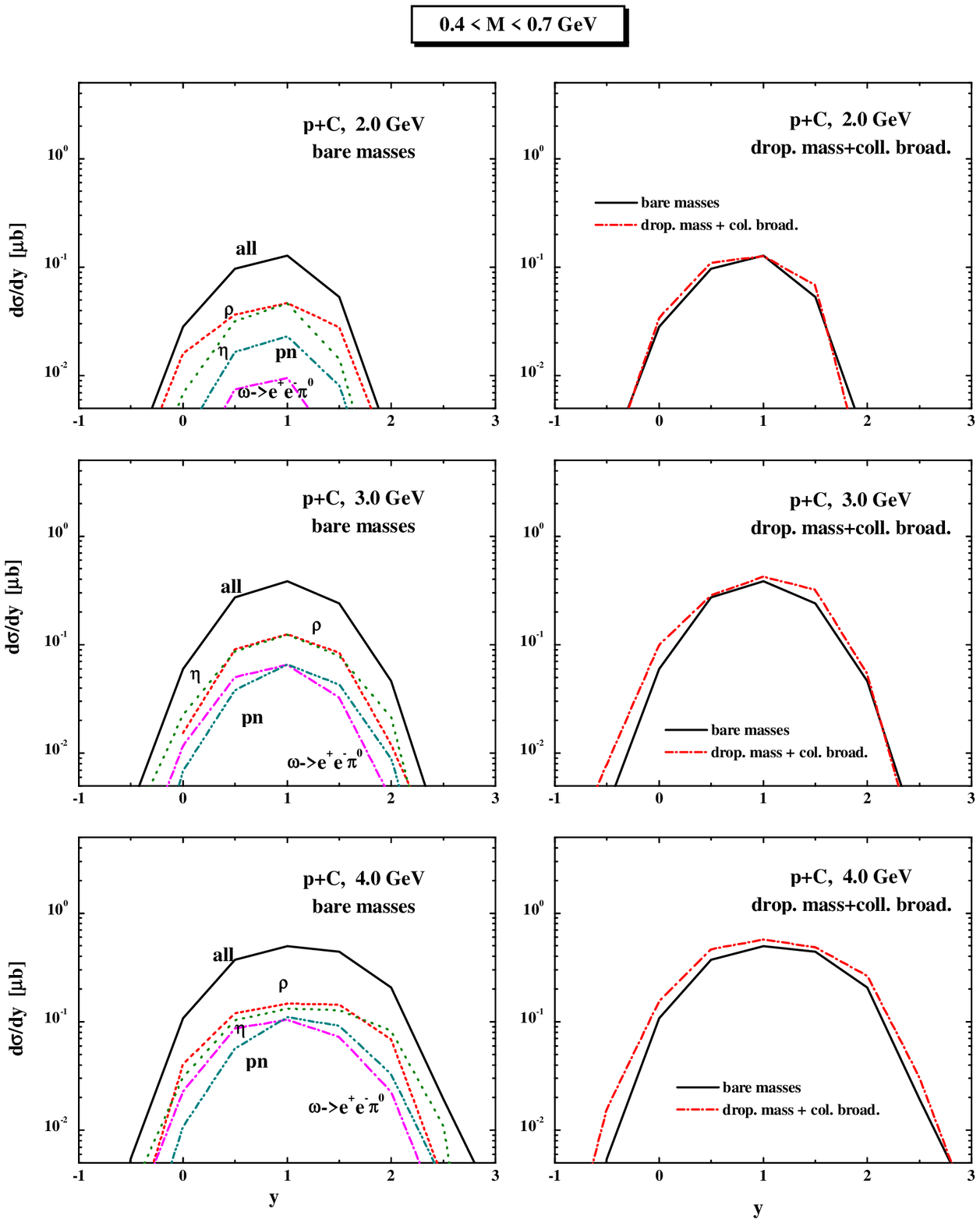,width=15cm}}
\caption{
The laboratory rapidity distributions $d\sigma/dy$ for
$p + C$ system at 1.0, 2.0, 3.0 and 4.0 GeV implying a cut in invariant
mass of $0.4 \le M \le 0.7$ GeV.  Left panel -- the individual
contributions for bare mass case, right panel - comparison of the bare
mass spectra (solid line) with the collisional broadening and dropping
mass scenario (dash-dotted lines). }
\label{Fig15pA}
\end{figure}

\begin{figure}[t]
\phantom{a}\vspace*{5mm}
\centerline{\psfig{figure=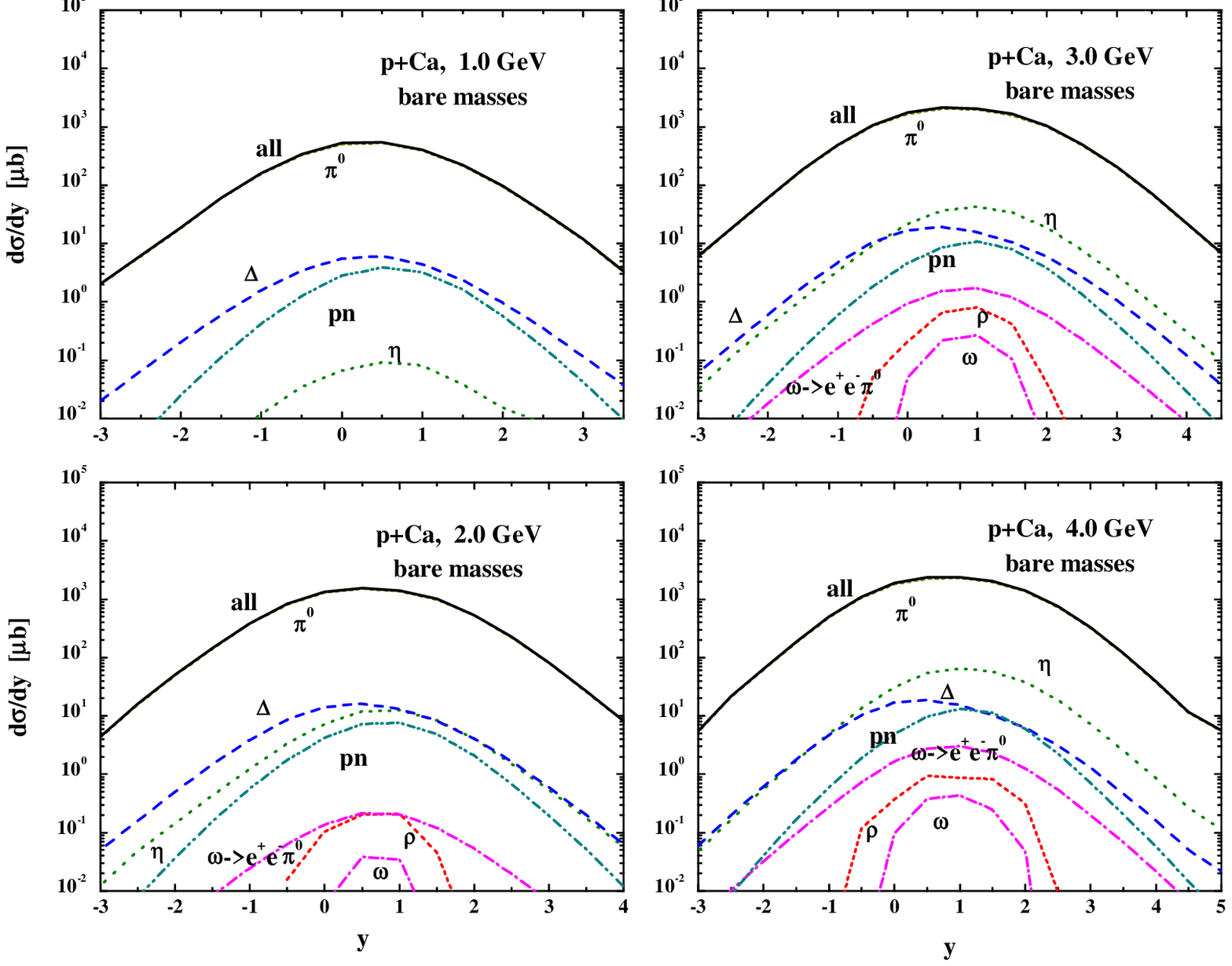,width=15cm}}
\vspace*{-8cm}
\caption{
The laboratory rapidity distributions $d\sigma/dy$ for the $p + Ca$ system
at 1.0, 2.0, 3.0 and 4.0 GeV. }
\label{Fig16pA}
\end{figure}

\begin{figure}[t]
\phantom{a}\vspace*{-20mm}
\centerline{\psfig{figure=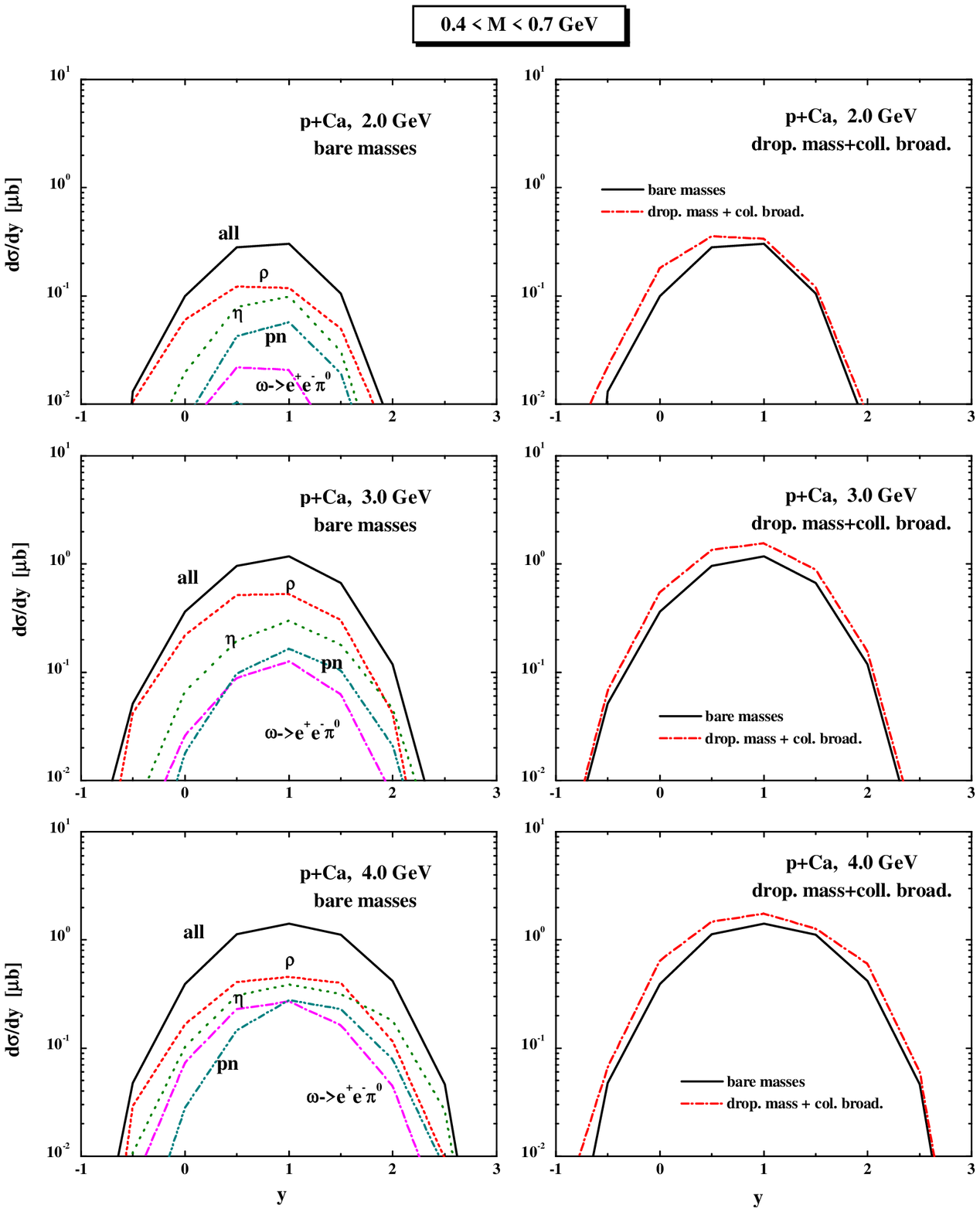,width=15cm}}
\caption{
The laboratory rapidity distributions $d\sigma/dy$ for
$p + Ca$ system at 1.0, 2.0, 3.0 and 4.0 GeV implying a cut in invariant
mass of $0.4 \le M \le 0.7$ GeV.  Left panel -- the individual
contributions for bare mass case, right panel - comparison of the bare
mass spectra (solid line) with the collisional broadening and dropping
mass scenario (dash-dotted lines). }
\label{Fig17pA}
\end{figure}

\begin{figure}[t]
\phantom{a}\vspace*{5mm}
\centerline{\psfig{figure=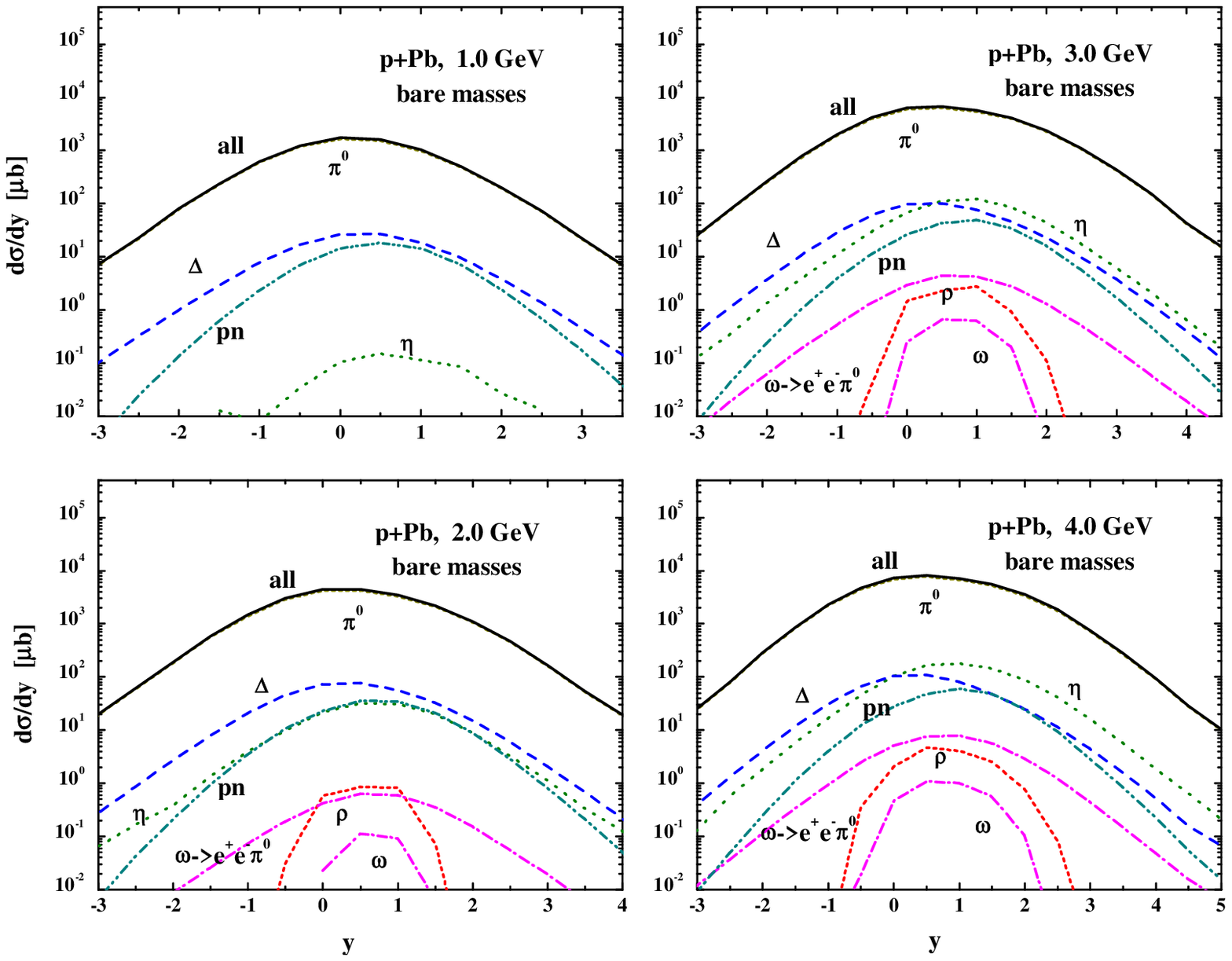,width=15cm}}
\vspace*{-8cm}
\caption{
The laboratory rapidity distributions $d\sigma/dy$ for the $p + Pb$ system
at 1.0, 2.0, 3.0 and 4.0 GeV. }
\label{Fig18pA}
\end{figure}

\begin{figure}[t]
\phantom{a}\vspace*{-20mm}
\centerline{\psfig{figure=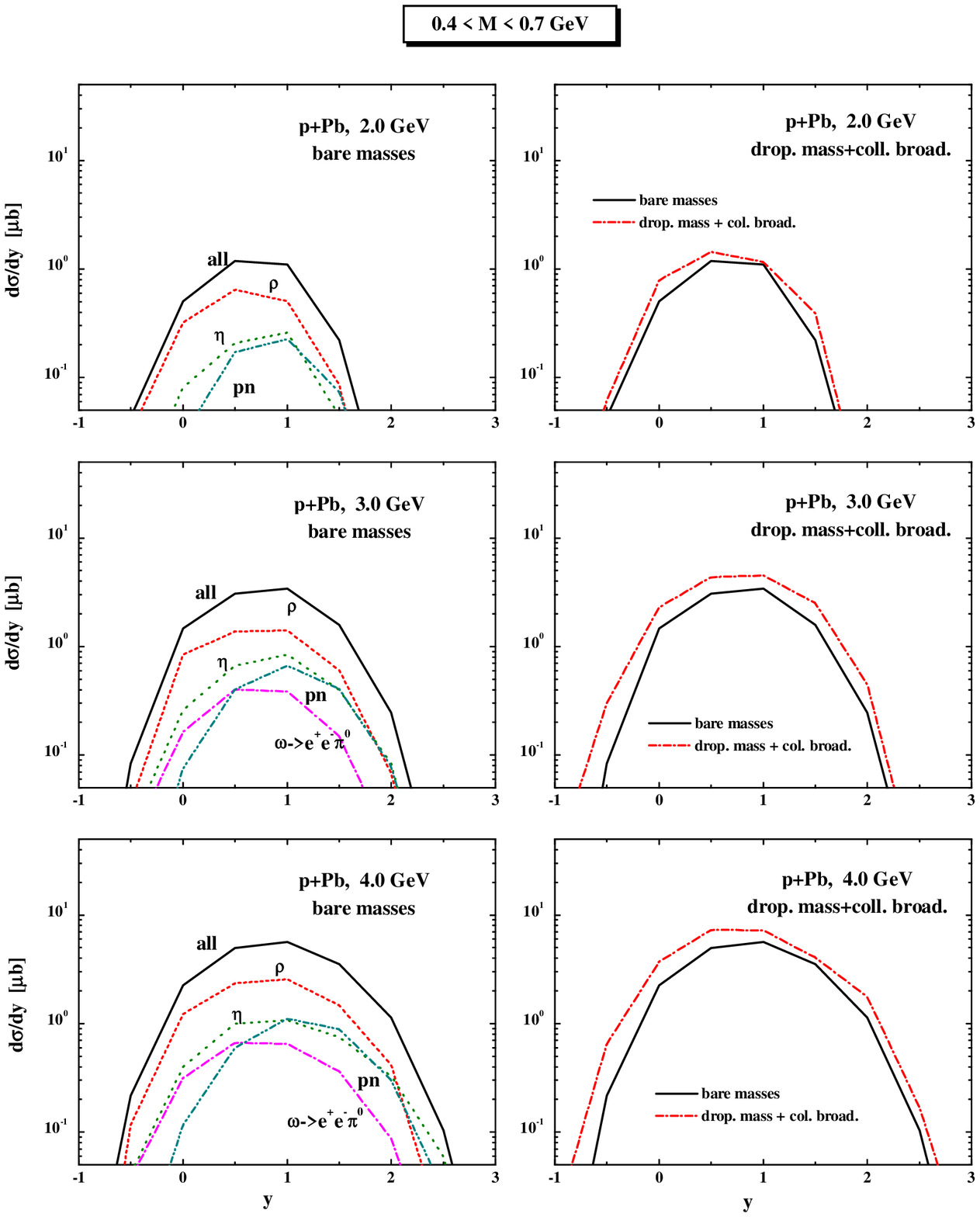,width=15cm}}
\caption{
The laboratory rapidity distributions $d\sigma/dy$ for
$p + Pb$ system at 1.0, 2.0, 3.0 and 4.0 GeV implying a cut in invariant
mass of $0.4 \le M \le 0.7$ GeV.  Left panel -- the individual
contributions for bare mass case, right panel - comparison of the bare
mass spectra (solid line) with the collisional broadening and dropping
mass scenario (dash-dotted lines). }
\label{Fig19pA}
\end{figure}

\begin{figure}[t]
\phantom{a}\vspace*{-10mm}
\centerline{\psfig{figure=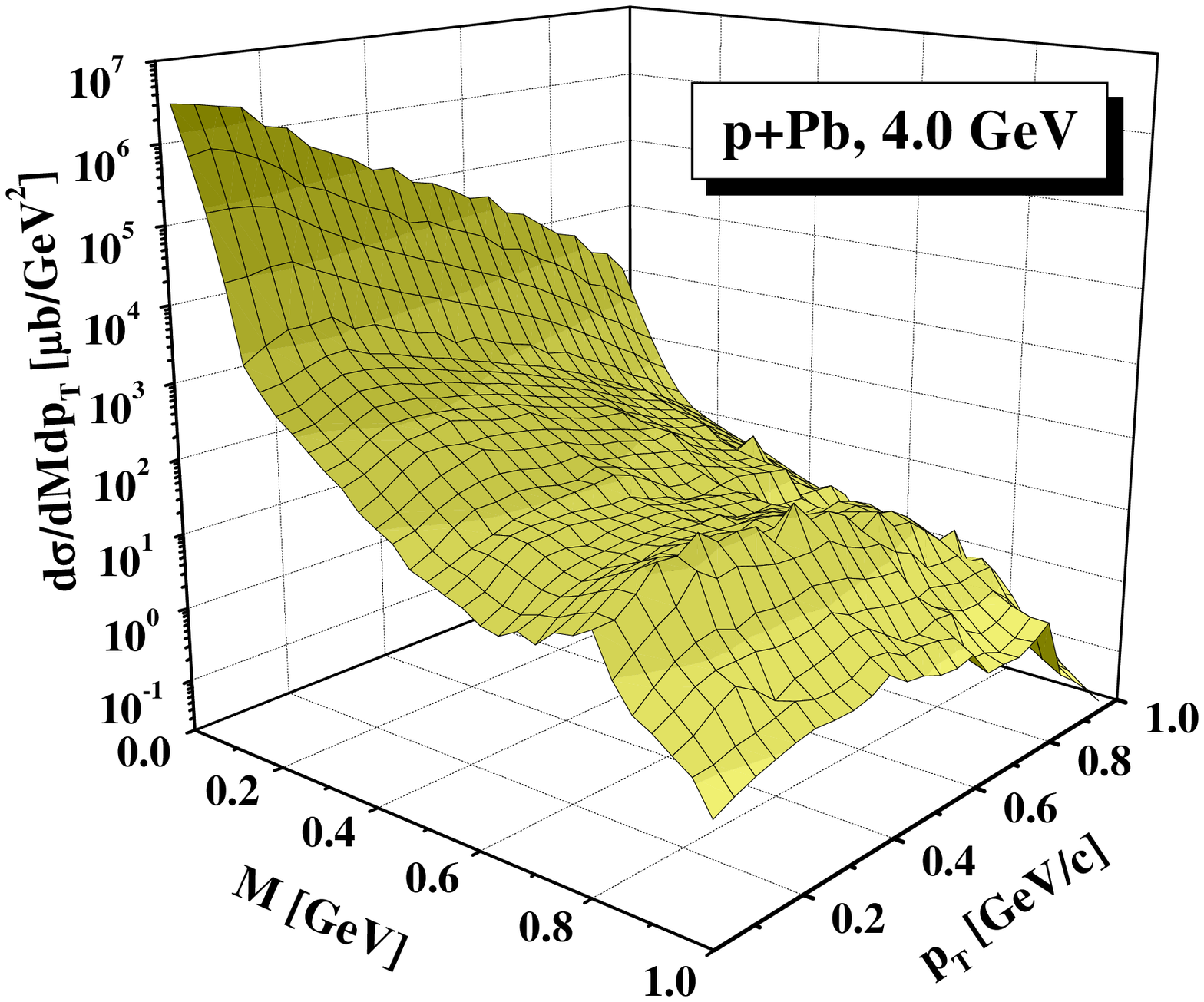,width=15cm}}
\caption{
The calculated double differential dilepton spectra $d\sigma/dMdp_T$
as a function of invariant mass $M$ and transverse momentum $p_T$
for $p + Pb$ collisions at 4.0 GeV.}
\label{Fig20pA}
\end{figure}

Finally, in Fig.~\ref{Fig20pA} the double differential dilepton
spectrum $d\sigma/dMdp_T$ is presented as a function of the invariant
mass $M$ and transverse momentum $p_T$ for $p + Pb$ collisions at 4.0
GeV calculated within the bare mass scenario.  At fixed $p_T$ one can
recognize the shape of the invariant mass spectra (cf. left panel of
Fig.~\ref{Fig6pA}) with a strong $\pi^0$ Dalitz decay branch at low $M$
as well as the contributions from $\eta$ Dalitz decay and the vector
meson ($\rho, \omega$) decays. At fixed $M$ the shape looks similar to
the one in Fig.~\ref{Fig6pA}. At low $M$ the exponential decrease
stems from the $\pi^0$ Dalitz decay, then the spectra become flatter
due to the contributions from $\eta$ Dalitz ($M\le 0.4$ GeV) and direct
$\rho$ decays. At $M\sim 0.78$ GeV the peak from the direct decay of
$\omega$ mesons is visible. Thus,  such type of 3-dimensional
experimental information (or even 4-dimensional including rapidity)
allows to select the contributions from different channels.

\section{Summary}

Within the framework of the coupled-channel (resonance) BUU model a
detailed nonperturbative study of dilepton production for $p+A$
reactions from 1--4 GeV has been performed employing a full off-shell
propagation of the vector mesons in line with Refs.
\cite{Cass_off1,Cass_off2}.  Different scenarios of in-medium
modifications of vector mesons, such as collisional broadening and
dropping vector meson masses, have been investigated and the
possibilities for an experimental observation of in-medium effects in
$p+A$ reactions has been discussed.

Dilepton spectra from $p+A$ reactions will be measured in future by the
HADES Collaboration at GSI Darmshtadt with high mass resolution and
good accuracy.  In this respect predictions for the dilepton invariant
mass spectra, transverse momentum and rapidity distributions for $p +
C$, $p + Ca$ and $p + Pb$ collisions from  1 to 4 GeV have been made
employing different in-medium scenarios. It has been  found that the
collisional broadening + 'dropping mass' scenario leads to an
enhancement of  the dilepton yield in the range $0.5 \leq M \leq 0.75$
GeV and to a reduction of the  $\omega$-peak, which is more pronounced
for heavy systems  (up to a factor 2 for $p + Pb$ at 3--4 GeV).

It has been indicated that proper cuts in invariant mass for transverse
momentum and rapidity spectra allow to select different dilepton
sources and to study, for example, the $\rho$ meson channel in more
detail.  However, an inclusion of in-medium effects predominantly leads
to an overall scaling in height of the spectra, but does not change the
slope of the $p_T$ and rapidity distributions very much.

It has been indicated, furthermore, that it might be very useful to
provide experimentally multi-dimensional information, e.g. double
differential dilepton spectra $d\sigma/dMdp_T$, in order to investigate
the individual contributions.

\acknowledgements
The author is grateful to U. Mosel, who initiated this work, for useful
suggestions and to W. Cassing for valuable discussions on the question
of off-shell transport theory.



\begin{references}
\bibitem{BrownRho}
    G.E. Brown and M. Rho, Phys. Rev. Lett. 66 (1991) 2720.
\bibitem{Shakin94}
    C.M. Shakin and W.-D. Sun, Phys. Rev. C 49 (1994) 1185.
\bibitem{Klingl96}
    F. Klingl and W. Weise, Nucl. Phys. A  606 (1996) 329;
    F. Klingl, N. Kaiser and W. Weise, Nucl. Phys. A  624 (1997) 527.
\bibitem{H&L92}
    T. Hatsuda and S. Lee,  Phys. Rev. C  46 (1992) R34.
\bibitem{Asakawa93}
    M. Asakawa and C.M. Ko, Phys. Rev. C  48 (1993) R526.
\bibitem{Leupold}
    S. Leupold, W. Peters and U. Mosel, Nucl. Phys. A  628 (1998) 311.
\bibitem{Herrmann}
        M. Herrmann, B. Friman, and W. N\"orenberg, Nucl. Phys. A  560
        (1993) 411.
\bibitem{asakawa}
        M. Asakawa, C. M. Ko, P. L\'evai, and X. J. Qiu, Phys. Rev.
        C  46 (1992) R1159.
\bibitem{Chanfray}
        G. Chanfray and P. Schuck, Nucl. Phys. A  545 (1992) 271c.
\bibitem{Rapp}
        R. Rapp, G. Chanfray, and J. Wambach, Phys. Rev. Lett.  76
        (1996) 368.
\bibitem{Friman}
        B. Friman and H. J. Pirner, Nucl. Phys. A  617 (1997) 496.
\bibitem{RappNPA}
        R. Rapp, G. Chanfray and J. Wambach, Nucl. Phys. A  617
        (1997) 472.
\bibitem{Peters}
        W. Peters, M. Post, H. Lenske, S. Leupold, and U. Mosel,
        Nucl. Phys. A  632 (1998) 109;
     M. Post, S. Leupold and U. Mosel, nucl-th/0008027.
\bibitem{CERES}
        G. Agakichiev et al., Phys. Rev. Lett.  75 (1995) 1272.
\bibitem{Ullrich}
        Th. Ullrich et al.,  Nucl. Phys. A  610 (1996) 317c;
        A. Drees, Nucl. Phys. A  610 (1996) 536c.
\bibitem{HELIOS}
        M. A. Mazzoni, Nucl. Phys. A  566 (1994) 95c;
        M. Masera, Nucl. Phys. A  590 (1995) 93c.
\bibitem{HELI2}
        T. {\AA}kesson et al., Z. Phys. C  68 (1995) 47.
\bibitem{Li}
        G. Q. Li, C. M. Ko, and G. E. Brown, Phys. Rev. Lett.  75
       (1995) 4007.
\bibitem{Li96}
        C. M. Ko, G. Q. Li, G. E. Brown, and H. Sorge,
        Nucl. Phys. A  610 (1996) 342c.
\bibitem{Cass95C}
        W. Cassing, W. Ehehalt, and C. M. Ko, Phys. Lett. B  363
       (1995) 35.
\bibitem{Cass96H}
        W. Cassing, W. Ehehalt, and I. Kralik,  Phys. Lett. B  377
      (1996) 5.
\bibitem{Brat97}
        E. L. Bratkovskaya and W. Cassing, Nucl. Phys. A  619 (1997) 413.
\bibitem{CBRep98}
        W. Cassing and E. L. Bratkovskaya,
        Phys. Rep.  308 (1999) 65.
\bibitem{Ernst}
        C. Ernst, S. A. Bass, M. Belkacem, H. St\"ocker, and W. Greiner,
        Phys. Rev. C 58 (1998) 447.
\bibitem{Ko93}
	 M. Asakawa and C. M. Ko, Nucl. Phys. A 560 (1993) 399.
\bibitem{Ko95}
	 G. Q. Li, C. M. Ko, and G. E. Brown, Nucl. Phys. A 606 (1996) 568.
\bibitem{Effe_piA}
     M. Effenberger, E. L. Bratkovskaya, W. Cassing and U. Mosel,
     Phys. Rev.  C 60 (1999) 027601.
\bibitem{CBRW97}
     W. Cassing, E. L. Bratkovskaya, R. Rapp, and J. Wambach,
     Phys. Rev. C  57 (1998) 916.
\bibitem{Effe99gam}
    M. Effenberger, E. Bratkovskaya and U. Mosel,
    Phys. Rev. C 60 (1999) 044614.
\bibitem{Wolf90}
     Gy. Wolf, G. Batko, W. Cassing, U. Mosel, K. Niita, and M. Sch\"afer,
     Nucl. Phys. A 517 (1990) 615;
     Gy. Wolf, W. Cassing and U. Mosel, Nucl. Phys. A 552 (1993) 549.
\bibitem{BCMas96}
        E. L. Bratkovskaya, W. Cassing and U. Mosel,
        Phys. Lett. B 376 (1996) 12.
\bibitem{EffePhD}
    M. Effenberger, Ph.D. Thesis, Univ. of Giessen, 1999;
    http://theorie.physik.uni-giessen.de/ftp.html.
\bibitem{Boresk96}
    K. G. Boreskov, J. Koch, L. A. Kondratyuk, and  M. I. Krivoruchenko,
    Phys. of Atomic Nuclei  59 (1996) 1908;
    K. G. Boreskov, L. A. Kondratyuk, M. I. Krivoruchenko, and J. Koch,
    Nucl. Phys. A 619 (1997) 295.
\bibitem{TeisZP97}
    S. Teis, W. Cassing, M. Effenberger, A. Hombach, U. Mosel,
    and Gy. Wolf, Z. Phys. A  356 (1997) 421;
    Z. Phys. A  359 (1997) 297.
\bibitem{Manley}
    D. M. Manley and E. M. Saleski, Phys. Rev. D  45 (1992) 4002.
\bibitem{FRITIOF}
    B. Anderson, G. Gustafson and Hong Pi, Z. Phys. C  57
    (1993) 485.
\bibitem{Ehehalt}
    W. Ehehalt and W. Cassing, Nucl. Phys. A  602 (1996) 449.
\bibitem{Bass}
    S. A. Bass et al., Prog. Part. Nucl. Phys. 42 (1998) 279;
    J. Phys. G 25 (1999) 1859.
\bibitem{BCM00SIS}
     E.L. Bratkovskaya, W. Cassing and U. Mosel,
     Nucl. Phys. A 686 (2001) 476.
\bibitem{Tuebingen}
    A. Faessler, C. Fuchs and M.I. Krivoruchenko,
    Phys. Rev. C 61 (2000) 035206.
\bibitem{GKC97}
    W. Cassing, Ye. S. Golubeva, A. S. Iljinov, and L. A. Kondratyuk,
    Phys. Lett. B  396 (1997) 26;
    Ye. S. Golubeva, L. A. Kondratyuk and W. Cassing,
    Nucl. Phys. A  625 (1997) 832.
\bibitem{Cass_off1}
    W. Cassing and S. Juchem, Nucl. Phys. A 665 (2000) 377.
\bibitem{Cass_off2}
    W. Cassing and S. Juchem, Nucl. Phys. A 672 (2000) 417.
\bibitem{Leupold_off}
    S. Leupold, Nucl. Phys. A 672 (2000) 475.
\bibitem{Kondr_rho}
    L. A. Kondratyuk, A. Sibirtsev, W. Cassing, Ye. S. Golubeva,
    and M. Effenberger, Phys. Rev. C 58 (1998) 1078.
\bibitem{Japen}
    K. Ozawa et al., nucl-ex/0011013.
\end{references}
\end{document}